\documentclass[aps,prx,reprint,superscriptaddress,floatfix]{revtex4-2}
\usepackage{amsmath,amssymb,amsfonts}
\usepackage{graphicx}
\usepackage{bm}
\usepackage{hyperref}
\hypersetup{hidelinks}
\usepackage{booktabs}

\newcommand{\CBdG}{C_{\rm BdG}}
\newcommand{\kB}{k_B}
\newcommand{\bk}{\bm{k}}
\newcommand{\dd}{\mathrm{d}}
\newcommand{\sgn}{\mathrm{sgn}}
\newcommand{\kmaj}{\frac{\pi^2\kB^2}{6h}}
\newcommand{\kdirac}{\frac{\pi^2\kB^2}{3h}}
\newcommand{\Gker}{\mathcal{G}}

\begin{document}

\title{Self-calibrating thermal interferometry of vortex parity in a two-dimensional chiral superconductor}

\author{Kumar Ghosh}
\email{jb.ghosh@outlook.com}
\affiliation{E.ON Digital Technology, Laatzener Str.\ 1,
             30539 Hannover, Germany}

\begin{abstract}
A chiral superconductor carries chiral Majorana modes along its boundary, and
the integer that counts them fixes everything that follows, yet that integer
has never been measured together with a local parity observable on one object.
Proximitized one-dimensional wires read fermion parity rapidly but diagnose
bulk topology through a separate protocol.  Here we show that a reconfigurable
domain wall between regions of opposite Chern number in an intrinsic
two-dimensional chiral superconductor performs both functions.  Opened to its
contacts the wall is a ballistic channel whose quantized thermal conductance
counts its Majorana modes; closed, the same wall is a Fabry--P\'erot resonator
whose spectrum shifts by half a level spacing when the parity of the enclosed
vortices changes, giving a two-level heat conductance.  We derive the exact
transmission, the elastic heat full counting statistics, and a theorem showing
that linear-response heat scattering of a fixed quadratic problem resolves
vortex parity but not the fusion channel of well-separated cores.  An outside
vortex hybridized with the wall is an intrinsic false positive; temperature,
geometry and a finite-bias mean--noise test separate it.
Rhombohedral-graphene parameters place submicron loops in the resolved regime
at millikelvin temperatures, where chiral-domain reconfiguration and noise
thermometry are both established.
\end{abstract}
\maketitle

%%==========================================================================
\section{Introduction}
\label{sec:intro}
%%==========================================================================

A superconductor whose gap winds in momentum space carries chiral Majorana
modes along its boundary and Majorana zero modes in its vortex cores, the
elementary ingredients of non-Abelian statistics and of hardware-protected
qubits~\cite{MooreRead1991,Read2000,Ivanov2001,Kitaev2003,Nayak2008}.  How many
such modes a given boundary carries is not a matter of degree: it is a signed
integer, the Bogoliubov--de Gennes (BdG) Chern number $\CBdG$, the topological
invariant that classifies the superconductor.  The practical obstacle is not
finding candidates but \emph{certifying} them.  A zero-energy feature can be a
Majorana mode anchored to a nonzero bulk invariant, or a fine-tuned trivial
Andreev level that mimics one, and distinguishing the two has proved
persistently difficult~\cite{Prada2020,Frolov2020,AguadoKouwenhoven2020}.

The leading experimental platform, a semiconductor nanowire proximitized by a
conventional superconductor, has advanced this problem furthest.  Interferometric
readout in indium-arsenide--aluminium devices achieves single-shot fermion-parity
measurement in microseconds with a reported $1\%$ assignment
error~\cite{MicrosoftMajorana2025}.  A structural feature of that architecture is
that bulk diagnosis and parity readout are carried out by different protocols on
different observables, so the link between the bulk invariant and the detected
parity is inferred rather than measured on the same object.  A one-dimensional
wire has no bulk boundary to interrogate.

A two-dimensional chiral superconductor does.  Bulk-boundary correspondence
ties $\CBdG$ directly to the number of chiral Majorana modes on any interface,
and that number is itself measurable, because each chiral Majorana channel
carries exactly half the thermal conductance
quantum~\cite{Read2000,Kane1997,Senthil1999}.  Quantized heat transport of
chiral edge modes has been resolved
experimentally~\cite{Jezouin2013,Banerjee2017,Banerjee2018}, and heat couples to
neutral modes that charge cannot see.  What has been missing is a design that
exploits the one feature a two-dimensional geometry offers and a wire cannot:
a propagation path whose shape can be changed.

Rhombohedral graphene now supplies exactly that.  Superconductivity emerges
there from an interaction-driven orbital ferromagnet without any proximity
layer~\cite{Han2025}; oppositely magnetized isospin domains have been imaged
inside the superconducting phase and reconfigured deterministically at
ultra-low current~\cite{Dutta2026}; multi-knob switching among spin-valley
polarized superconducting states has been demonstrated in hexalayer
devices~\cite{Hua2026}; and microscopic theory identifies valley polarization
as selecting the $p$-wave chirality, with a Lifshitz transition separating
topological from trivial pairing~\cite{Geier2025}.  Floating-contact
Johnson-noise thermometry, meanwhile, already operates on
hexagonal-boron-nitride-encapsulated graphene at millikelvin
temperatures~\cite{Srivastav2019,Srivastav2022,Waissman2021}.  The ingredients
for a programmable, heat-read Majorana interferometer are therefore in place
in a single material system.

Here we show what such a device measures.  A closed domain wall between
$\CBdG=+C$ and $-C$ regions, coupled to two floating thermal contacts, operates
in two modes on the same physical object.  With the contacts open it is a
ballistic channel whose quantized conductance counts its Majorana modes and
thereby calibrates the bulk invariant.  With the contacts closed it is a
Fabry--P\'erot resonator whose spectrum is set by the boundary condition of the
Majorana field, which the parity of the enclosed vortices switches between
antiperiodic and periodic, shifting the resonance comb by half a level spacing
and producing a two-level thermal conductance.  We derive the exact
transmission and its heat full counting statistics; we prove that
linear-response heat scattering of a fixed quadratic problem resolves the
vortex parity but not the Ising fusion channel of well-separated cores, which
delimits the claim precisely; we identify edge-vortex hybridization as an
intrinsic false positive and give temperature, geometry and mean--noise
discriminators against it; and we show that representative
rhombohedral-graphene parameters place submicron loops in the resolved regime
at temperatures already reached by existing thermometry.

Throughout, Ref.~\cite{Ghosh2026THT} supplies two inputs used here: the occupied-vortex rule fixing $\CBdG$ from the
pairing-vortex winding enclosed by the occupied Fermi sea, and the gappedness criterion $\delta_{\rm BdG}>0$.  In the finite-$\bm Q$ decomposition the
$\xi_a(\bk)\tau_0$ term is eigenvector-inert, whereas warping and finite
$\bm Q$ can also modify the $\tau_z$ component.  Consequently topology is
assumed unchanged only along a continuous deformation that remains gapped,
as checked explicitly in Ref.~\cite{Ghosh2026THT}.

%%==========================================================================
\section{Physical principle}
\label{sec:formulation}
%%==========================================================================

\subsection{Vortex parity and edge spin structure}

Let $\Sigma$ be a two-dimensional superconductor in symmetry class D with
Chern number $\CBdG$.  For $\CBdG$ odd, an $n_v=1$ Abrikosov vortex binds a
single unpaired Majorana zero mode~\cite{Volovik1999,Read2000,Ivanov2001} and
is an Ising $\sigma$ defect.  The underlying non-Abelian theory obeys
\begin{equation}
  \sigma\times\sigma = 1 + \psi ,
\label{eq:fusion}
\end{equation}
but the observable constructed here is deliberately more limited than a full
fusion-channel measurement.  It is the vortex-parity-induced change of the
closed-edge spin structure,
\begin{equation}
  n_v\ {\rm even}:\ {\rm NS},\qquad
  n_v\ {\rm odd}:\ {\rm R}.
\label{eq:vortex-sector}
\end{equation}
The two sectors differ by a half-level shift and the Ramond sector contains a
zero mode.  In Ising-CFT language the odd-vortex sector carries topological
charge $\sigma$, whereas an even number of vortices can fuse to either $1$ or
$\psi$.  Both even-vortex outcomes have Neveu--Schwarz boundary conditions,
so the single-particle spectrum used here does not distinguish $1$ from
$\psi$.  This distinction is essential: the interferometer is a bulk-calibrated
vortex-parity probe, not a fusion-rule measurement.

\subsection{Thermal readout of a closed edge}

The construction combines three ingredients that are compatible with the same
rhombohedral-graphene device.

\emph{(i) A neutral probe.}  Chiral Majorana modes carry no charge, so an
electrical interferometer must first fuse them into a Dirac
mode~\cite{FuKane2009}, which reintroduces sensitivity to charge disorder and
to the Aharonov-Bohm phase.  Heat couples to every mode regardless of charge,
and a chiral Majorana channel carries exactly half the Dirac thermal
conductance quantum,
\begin{equation}
  \frac{K}{T}\bigg|_{\rm one\ Majorana}
  = \kmaj \simeq 4.732\times10^{-13}\ \frac{\rm W}{\rm K^2}.
\label{eq:majorana-quantum}
\end{equation}

\emph{(ii) A closed, reconfigurable path.}  The $Z_2$ boundary-condition
switch is defined only for a closed circuit enclosing the defect.  A locally
pinned domain wall can close on itself and can, in principle, be reshaped
between measurements.  Dutta et al. establish deterministic domain
reconfiguration, while arbitrary closed-loop writing remains an experimental
requirement~\cite{Dutta2026}.

\emph{(iii) A resolvable finite-size scale.}  The relevant finite-size scale is
accessible at dilution-refrigerator temperatures because the chiral Majorana edge
velocity of a weak-pairing chiral superconductor is
$v=\Delta_0/\hbar k_F$, of order $10^3$~m/s in rhombohedral graphene, three
orders of magnitude below the graphene Fermi velocity.  The basic finite-size scale $\hbar v/L$ is therefore of order $10$~mK for a
micron-scale loop; the Neveu--Schwarz excitation gap is $\pi\hbar v/L$.
Both scales are accessible at temperatures already reached by graphene noise
thermometry.

\subsection{From bulk invariant to resonant heat transport}

Bulk-boundary correspondence first fixes the number of co-propagating
Majorana branches on a $+C/-C$ wall.  Enclosed vortex parity then selects the
Neveu--Schwarz or Ramond spin structure and shifts the finite-size spectrum.
Two point contacts convert this spectral shift into a transmission resonance,
whose multiple-traversal series can be summed in closed form; the Majorana
Landauer formula then gives the finite-temperature heat conductance.  The
remaining material scales, principally the edge-mode velocity and the
perpendicular critical field, determine whether the resonance and vortex
controls can be accessed in the same device.

%%==========================================================================
\section{The rewritable wall as a closed chiral Majorana edge}
\label{sec:wall}
%%==========================================================================

\subsection{Mode counting}
\label{subsec:counting}

Consider a region $\Omega$ of reversed chirality written inside a chiral
superconductor, so that $\CBdG=+C$ outside and $-C$ inside.  Bulk-boundary
correspondence for symmetry class D gives, at the interface,
\begin{equation}
  N = |C_{\rm out}-C_{\rm in}| = 2|C|
\label{eq:mode-count}
\end{equation}
co-propagating chiral Majorana channels, circulating with a handedness fixed
by $\sgn(C_{\rm out}-C_{\rm in})$~\cite{Read2000,StoneRoy2004,TeoKane2010}.
Each channel is a real chiral fermion of central charge $c=1/2$; the total
chiral central charge of the wall is therefore $|C|$.  For $|C|=1$ the wall
contains two co-propagating Majorana branches, not a single Ising edge.

Two features of this counting matter for what follows.  First, it is
independent of the microscopic profile of the wall: only the difference of
bulk invariants enters.  Second, the construction requires the adjacent bulk
domains to remain in the same gapped topological phases under microscopic
deformations.  In the finite-$\bm Q$ formulation of Ref.~\cite{Ghosh2026THT},
$\xi_a(\bk)\tau_0$ does not change eigenvectors, whereas warping and finite
$\bm Q$ can also modify the spectrally active $\xi_s\tau_z$ component.  The
safe criterion is therefore a continuous gapped deformation.  The condition
\begin{equation}
  \delta_{\rm BdG}=\min_{\bk}\Bigl[\sqrt{\xi_s^2+|\Delta_{\bk}|^2}-|\xi_a|\Bigr]>0
\label{eq:gapped}
\end{equation}
bounds the entire construction: where it fails, the adjacent bulk is gapless,
the wall is not isolated, and no statement below applies.

For the remainder we take $|C|=1$, so the wall carries $N=2$ Majorana
branches.  Equations below first use the independent-channel limit as an
analytically transparent baseline.  Class D does not protect branch
independence, and recent microscopic work on rhombohedral-graphene domain
walls shows that intervalley mixing can materially affect transport across
opposite-chirality regions~\cite{Phong2026}.  A symmetry-allowed static
$SO(2)$ mixing of the co-propagating branches is therefore included explicitly
in Appendix~\ref{app:holonomy}; it renormalizes the resonant contrast without
changing the open-contact plateau.

\subsection{Enclosed charge and the spin structure}
\label{subsec:sectors}

A chiral Majorana mode on a closed contour of circumference $L$ admits two
spin structures.  In a superconducting realization the parity of the enclosed
unit-vortex number fixes whether the Majorana field is antiperiodic or periodic
around the contour~\cite{DiFrancesco1997,Nayak2008,Kitaev2006,Alicea2012}.

The mechanism is the standard branch-cut argument.  A shift of the
superconducting phase by $2\pi$ is invisible to Cooper pairs but reverses the
sign of any unpaired fermion, so an $h/2e$ vortex may be represented by a
branch cut emanating from its core, and a Majorana operator transported across
that cut acquires a minus sign~\cite{Ivanov2001,Alicea2012}.  A contour
enclosing an odd number of vortices therefore crosses an odd number of cuts
and converts antiperiodic into periodic boundary conditions.  Alicea reaches
exactly this conclusion for the two-vortex geometry, where the inner edges
carry periodic boundary conditions and host zero modes while the outer edge
remains antiperiodic and retains a finite-size gap~\cite{Alicea2012}.  What
follows applies that same rule to a contour whose shape is set by a gate
rather than by a sample boundary.

Transporting the Majorana operator once around the loop therefore returns it
multiplied by a $Z_2$ boundary-condition sign.  We use the convention
\begin{equation}
  \sigma = (-1)^{n_v+1},
\label{eq:z2}
\end{equation}
so that $n_v$ even gives $\sigma=-1$ (Neveu--Schwarz) and $n_v$ odd gives
$\sigma=+1$ (Ramond).  This sign diagnoses the parity of the number of
Majorana-carrying vortices.  It should not be confused with the even-vortex
fusion label: both total charges $1$ and $\psi$ have Neveu--Schwarz boundary
conditions and therefore the same single-particle resonance comb.

The consequence for the spectrum is immediate.  With $k$ the momentum along
the wall and $\varepsilon = \hbar v k$,
\begin{equation}
  \varepsilon_n =
  \begin{cases}
    \dfrac{2\pi\hbar v}{L}\,n, & \sigma=+1 \quad \text{(Ramond)},\\[6pt]
    \dfrac{2\pi\hbar v}{L}\left(n+\tfrac12\right), & \sigma=-1
      \quad \text{(Neveu-Schwarz)},
  \end{cases}
\label{eq:loop-spectrum}
\end{equation}
with $n\in\mathbb{Z}$.  The Ramond comb contains $\varepsilon=0$; the
Neveu--Schwarz comb is offset by half a period and has a minimum excitation
energy $\pi\hbar v/L$.  The Ramond zero mode is the finite-size edge
manifestation of the odd-vortex spin structure.  Global fermion-parity
constraints determine how this edge zero mode combines with vortex-core zero
modes, but that many-body occupation is not resolved by the single-particle
thermal spectrum calculated here.

\begin{figure*}[t]
  \centering
  \includegraphics[width=0.95\textwidth]{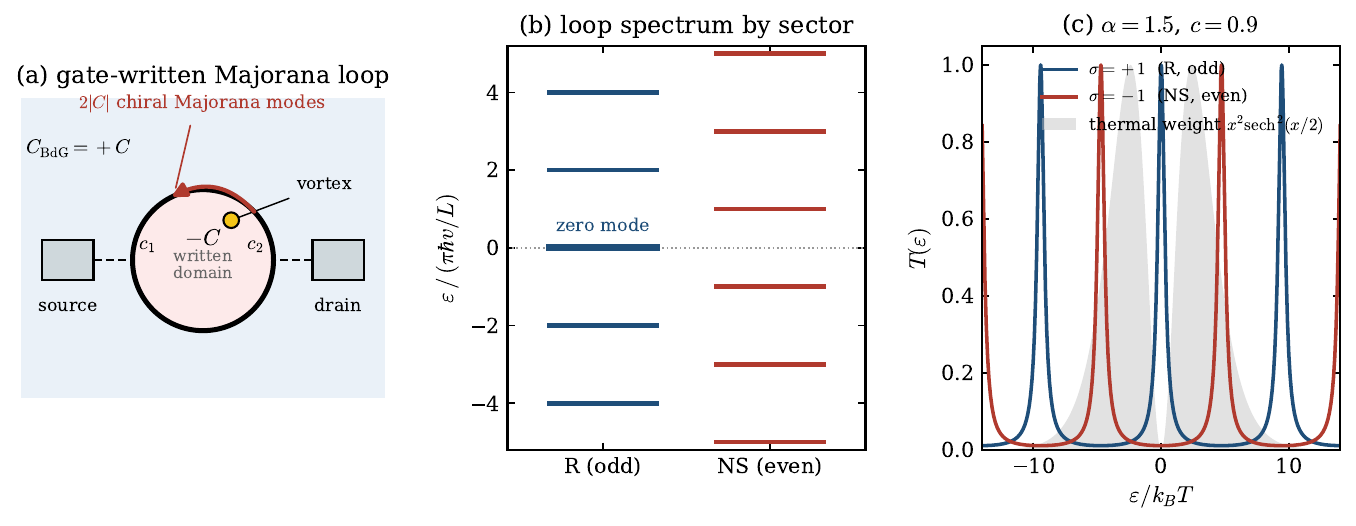}
  \caption{The rewritable Majorana loop and its two vortex-parity sectors.
    (a)~A locally pinned region of reversed chirality is formed inside a chiral
    superconductor; the closed wall carries $2|C|$ co-propagating chiral
    Majorana modes and is coupled to floating source and drain contacts
    through two point contacts with loop-return amplitudes $c_1,c_2$.  The
    enclosed vortex is an Ising $\sigma$ defect.
    (b)~Single-particle spectrum of one Majorana channel on the loop,
    Eq.~\eqref{eq:loop-spectrum}.  Odd enclosed parity selects the Ramond
    sector, which contains a zero mode; even parity selects Neveu-Schwarz,
    whose combs are offset by half a period.
    (c)~Transmission $T(\varepsilon)$, Eq.~\eqref{eq:transmission}, for the two
    sectors at $\alpha=1.5$ and $c=0.9$, with the thermal weight
    $x^2\,\mathrm{sech}^2(x/2)$ shaded.  Because that weight vanishes at
    $\varepsilon=0$, the Ramond zero mode carries heat only through its finite
    width, which is what makes the two sectors thermally distinguishable.}
  \label{fig:concept}
\end{figure*}

This is the interferometric principle: enclosed vortex parity is
converted into a half-level shift of a closed chiral-Majorana resonance comb.
What remains is to show that a heat measurement can read that shift.
Figure~\ref{fig:concept}(a,b) shows the geometry and the two spectra.

\subsection{Material scales}
\label{subsec:scales}

For a weak-pairing chiral superconductor with $\Delta_{\bk}=\Delta_0 k/k_F$, the
chiral edge mode disperses linearly with velocity~\cite{Read2000,StoneRoy2004}
\begin{equation}
  v = \frac{\Delta_0}{\hbar k_F}.
\label{eq:edge-velocity}
\end{equation}
This is not the Fermi velocity.  For a general multiband chiral state,
$v$ should be understood as the actual slope of the domain-wall Majorana
dispersion; Eq.~\eqref{eq:edge-velocity} is the weak-pairing benchmark used
for the scale estimates below.  For the numerical scale estimates we take $T_c=300$~mK from Han
et al.~\cite{Han2025} and the representative value
$k_F=0.29$~nm$^{-1}$ quoted by Okounkova et al. from their degeneracy-two
Landau-fan analysis in octalayer graphene~\cite{Okounkova2026}.  These inputs
are not a fit to one device: the comparatively large $k_F$ is used as a
representative RHG momentum scale and gives a lower, hence conservative, edge
velocity at fixed gap.  Gap ratios $2\Delta_0/\kB T_c$ between $10$ and
$30$ then give $\Delta_0=0.13$--$0.39$~meV and
\begin{equation}
  v \simeq 0.7\text{--}2.0\times10^{3}\ \mathrm{m/s},
\label{eq:v-numbers}
\end{equation}
three orders of magnitude below the graphene Fermi velocity.  Two derived scales are relevant.

The reduced finite-size scale is
$\hbar v/(L\kB) \simeq 5$--$16$~mK for $L=1\,\mu$m.  Equivalently, the
lowest Neveu--Schwarz excitation lies at $\pi\hbar v/L$, corresponding to
approximately $16$--$49$~mK.  Thus, at the $12$~mK base temperature of
Ref.~\cite{Srivastav2019}, the dimensionless ratio
\begin{equation}
  \alpha \equiv \frac{\hbar v}{L\,\kB T}
\label{eq:alpha}
\end{equation}
is of order unity.  These parameters place the representative device near the crossover where the
sectors are resolved but the conductance is not yet exponentially small, the
regime of largest contrast.

The Majorana thermal wavelength
$\lambda_T = 2\pi\hbar v/\kB T$ is $2.7$--$8.1\,\mu$m at $12$~mK, several
times longer than a submicron loop.  This estimate does not determine the
inelastic coherence length, however, because wall roughness and equilibration
of the RHG domain-wall modes have not yet been characterized.

Fu and Kane noted that the electrical neutrality of a chiral Majorana mode
suppresses direct electromagnetic coupling and that the lowest allowed local
self-interaction of a single chiral Majorana field,
$\gamma\,\partial_x\gamma\,\partial_x^2\gamma\,\partial_x^3\gamma$, contains
six spatial derivatives and is strongly irrelevant at low
temperature~\cite{FuKane2009}.  These observations motivate the coherent-loop
regime assumed below, but do not by themselves establish it for rhombohedral
graphene.  We therefore treat an inelastic/equilibration length exceeding the
loop perimeter as an explicit device requirement rather than a derived
material property.  The open-contact plateau remains useful even if this stronger coherence
condition is not met, whereas parity-sensitive resonances require coherent
propagation around the loop.

%%==========================================================================
\section{Scattering theory and thermal response}
\label{sec:scattering}
%%==========================================================================

\subsection{Transmission through the loop}
\label{subsec:transmission}

Couple the loop to two floating reservoirs through point contacts at positions
separating the wall into arcs $L_1$ and $L_2$, with $L=L_1+L_2$.  Particle-hole
symmetry implies $S(\varepsilon)=S^*(-\varepsilon)$, so the Majorana
scattering matrix is real at zero energy.  In the leading low-energy,
energy-independent contact approximation it may therefore be taken as a real
orthogonal $2\times2$ matrix, leaving one mixing angle.  Writing
$c_i=\cos\theta_i$
for the amplitude to remain on the loop and $s_i=\sin\theta_i$ for the
amplitude to enter the reservoir,
\begin{equation}
  \begin{pmatrix}\text{loop}_{\rm out}\\ \text{lead}_{\rm out}\end{pmatrix}
  =\begin{pmatrix} c_i & s_i\\ -s_i & c_i\end{pmatrix}
  \begin{pmatrix}\text{loop}_{\rm in}\\ \text{lead}_{\rm in}\end{pmatrix},
  \qquad c_i^2+s_i^2=1 .
\label{eq:contact-S}
\end{equation}
The one-angle contact is the minimal low-energy model used for the closed-form
resummation below.  Analytic energy dependence of the contact scattering adds
nonuniversal corrections that are small when the contact scale exceeds
$\kB T$ and $\hbar v/L$.

Propagation along an arc of length $\ell$ multiplies the amplitude by
$e^{i\varepsilon\ell/\hbar v}$, and one complete traversal of the loop carries
in addition the $Z_2$ sign of Eq.~\eqref{eq:z2}.  Summing the geometric series
over the number of traversals gives the transmission amplitude from reservoir
$1$ to reservoir $2$,
\begin{equation}
  t_{21}(\varepsilon)
  = \frac{s_1 s_2\,e^{i\varepsilon L_1/\hbar v}}
         {1-c_1c_2\,\sigma\,e^{i\varepsilon L/\hbar v}},
\label{eq:t-amplitude}
\end{equation}
and hence the transmission probability
\begin{equation}
  \boxed{\;
  T(\varepsilon)
  = \frac{(1-c_1^2)(1-c_2^2)}
         {1+c_1^2c_2^2-2c_1c_2\,\sigma\cos\!\left(\varepsilon L/\hbar v\right)} .\;}
\label{eq:transmission}
\end{equation}
Within this minimal single-branch contact model only the total perimeter enters,
not the individual arcs.  Additional contact phases or spatially nonuniform
multibranch mixing can introduce nonuniversal geometric dependence without
altering the vortex-parity spin-structure constraint.

Equation~\eqref{eq:transmission} has unit peaks wherever
$\cos(\varepsilon L/\hbar v)=\sigma$, reproducing the spectra of
Eq.~\eqref{eq:loop-spectrum} as resonances, with width set by $c_1c_2$.  For
transparent contacts, $c_1=c_2=0$, it collapses to $T\equiv1$ for all
$\varepsilon$ and all $\sigma$: the loop becomes an ideal ballistic channel and
the sector information is lost.  This limit is not a defect of the scheme but
its calibration point, as shown below.

\paragraph{Relation to two-arm Majorana interferometers.}
The established Majorana interferometers are two-arm devices in which an
incident electron splits into a pair of Majorana modes that traverse arms of
lengths $L_b$ and $L_c$ and recombine, so that the relevant phase is
$\varepsilon\,\delta L/\hbar v + n_v\pi$ with
$\delta L=L_b-L_c$~\cite{FuKane2009,Akhmerov2009}.  That construction has a
property the present one does not: at $\delta L\to0$ the dynamical phase
disappears and, in the ideal converter model, the $Z_2$ readout becomes
temperature independent,
the conductance taking the values $0$ or $2e^2/h$ according to the parity of
$n_v$~\cite{FuKane2009,Akhmerov2009,Alicea2012}.  It also has a property that
eliminates parity contrast in the corresponding heat measurement: the parity determines whether the
outgoing quasiparticle is an electron or a hole, and every incident excitation
reaches the drain either way, so the \emph{heat} transmitted is unity in both
sectors.

A closed resonator inverts both statements.  Because the interfering paths
differ by whole traversals, only the perimeter $L$ enters and the visibility is
controlled by $\alpha$ rather than by an arm asymmetry, which is the price paid
in Sec.~\ref{subsec:contrast}.  In exchange, the parity now moves the
\emph{spectrum} rather than the outgoing particle species, and a spectrum is
exactly what a heat current is sensitive to.  The two geometries are therefore
complementary probes of the same $Z_2$ quantum number.  As discussed in
Sec.~\ref{subsec:charge}, combining them on one rhombohedral-graphene device
would additionally require a genuine charged chiral channel and a
Dirac--Majorana converter.

\subsection{Thermal conductance}
\label{subsec:conductance}

The linear-response heat conductance of a chiral Majorana channel with
transmission $T(\varepsilon)$ follows from the transport thermal
coefficient~\cite{QinNiuShi2011,Ghosh2026THT}, with the factor $1/2$ that
distinguishes a real from a complex fermion,
\begin{equation}
  K = \frac{1}{2}\,\frac{1}{2\pi\hbar T}
      \int\!\dd\varepsilon\;\varepsilon^2
      \left(-\frac{\partial f}{\partial\varepsilon}\right)T(\varepsilon).
\label{eq:K-landauer}
\end{equation}
Normalizing to the Majorana quantum of Eq.~\eqref{eq:majorana-quantum} and
writing $x=\varepsilon/\kB T$,
\begin{equation}
  \frac{K}{\pi^2\kB^2T/6h}
  = \frac{3}{\pi^2}\int_{-\infty}^{\infty}\!\dd x\;
    \frac{x^2}{4}\,\mathrm{sech}^2\!\left(\frac{x}{2}\right) T(x),
\label{eq:K-dimensionless}
\end{equation}
with $T(x)$ given by Eq.~\eqref{eq:transmission} at
$\varepsilon L/\hbar v = x/\alpha$.  The normalization is fixed by
$\int\dd x\,(x^2/4)\,\mathrm{sech}^2(x/2)=\pi^2/3$, so that transparent
contacts return exactly one Majorana quantum per channel.

Two limits connect this expression to Ref.~\cite{Ghosh2026THT}.  For $T\equiv1$,
Eq.~\eqref{eq:K-dimensionless} gives the plateau
\begin{equation}
  \frac{K_{\rm DW}}{T}\bigg|_{c_i\to0}
  = 2|C|\,\kmaj = |C|\,\kdirac,
\label{eq:plateau-limit}
\end{equation}
which is the quantized domain-wall conductance predicted in
Ref.~\cite{Ghosh2026THT}: for $|C|=1$ the wall carries one full Dirac thermal
quantum although it is built from two Majorana channels.  The Fermi thermal weight in Eq.~\eqref{eq:K-dimensionless} is also the one
appearing in the bulk curvature-resolved kernel of Ref.~\cite{Ghosh2026THT},
\begin{equation}
  \Gker(x)=\frac{12}{\pi^2}\int_0^x u^2\,\mathrm{sech}^2u\;\dd u
\label{eq:Gkernel}
\end{equation}
whose closed form in terms of the dilogarithm and asymptotics
$\Gker\to(4/\pi^2)x^3$ and $\Gker\to1$ are quoted for comparison.  The two
objects should not be conflated: the transparent resonator has
$T(\varepsilon)\equiv1$ and therefore gives the quantized value directly,
not $\Gker(x)$.

The essential structural point is visible in Fig.~\ref{fig:concept}(c).  The
thermal weight $\varepsilon^2(-\partial f/\partial\varepsilon)$ vanishes
quadratically at $\varepsilon=0$.  A zero-energy quasiparticle carries no
heat, so the Ramond zero mode contributes only through the finite width of its
resonance.  The two sectors are therefore distinguished not by the presence or
absence of a conducting level but by \emph{where the first heat-carrying level
sits}: at $2\pi\hbar v/L$ in the Ramond sector and at $\pi\hbar v/L$ in
Neveu-Schwarz.  This is a thermal even-odd effect, topological in origin, and
it is what the next section quantifies.

%%==========================================================================
\section{Results}
\label{sec:results}
%%==========================================================================

\subsection{Fusion blindness of quadratic heat scattering}
\label{subsec:nogo}

The thermal observable has sharply delimited information content.

Let the loop enclose two vortices with core Majorana modes $\gamma_1,\gamma_2$.
Their Ising fusion channel is the eigenvalue of $\mathcal{P}=i\gamma_1\gamma_2$,
with $\mathcal{P}=+1$ for the vacuum channel $1$ and $-1$ for the fermion
channel $\psi$.  Write the full Bogoliubov--de Gennes problem in the Majorana
basis $\chi=\{\eta_j,\gamma_1,\gamma_2\}$, with $\eta_j$ the discretized wall
field,
\begin{equation}
  H=\frac{i}{2}\sum_{ab}A_{ab}\,\chi_a\chi_b ,
  \qquad A=-A^{T}\in\mathbb{R}^{M\times M}.
\label{eq:bdg-majorana}
\end{equation}
Every quantity entering linear-response heat transport, the scattering matrix
$S(\varepsilon)$, the transmission $T(\varepsilon)$, and hence $K$, is a
functional of $A$ alone.

\paragraph{No-go.}
\emph{For well-separated vortex cores, the direct overlap
$A_{\gamma_1\gamma_2}$ is exponentially small and the $1$ and $\psi$ states
form the degenerate fusion manifold.  Within a fixed quadratic BdG scattering
problem, changing the occupation label $\mathcal{P}=i\gamma_1\gamma_2$ does
not change $A$ and therefore cannot change the linear-response thermal
conductance.  The enclosed-vortex dependence retained by the loop is the
boundary-condition sign $\sigma=(-1)^{n_v+1}$.}

The underlying reason is more general than the vanishing overlap.  For a
fixed quadratic Hamiltonian, $A$ determines the single-particle scattering
matrix, whereas $\mathcal{P}$ labels the occupation of the many-body zero-mode
manifold and is not an independent parameter of $A$.  A direct overlap term
$i\epsilon\gamma_1\gamma_2$ would split that manifold, but it would represent a
Hamiltonian perturbation rather than make the scattering matrix conditional on
which fusion state was prepared.  Edge-vortex hybridization and fermion
tunneling between wall branches likewise modify $A$ without promoting the
fusion occupation to a scattering parameter.  Appendix~\ref{app:nogo}
exhibits the separated-core matrix explicitly.

The consequence is a sharp statement about this class of quadratic
linear-response scattering measurements.  Making transport conditional on the
fusion occupation therefore requires physics beyond a fixed quadratic
scattering problem.  Standard routes include
non-Gaussian $\sigma$ transport, such as vortex or edge-vortex tunneling, and
charging-energy constraints on a floating island.  Representative fusion-readout
schemes implement one of these mechanisms.  Klocke
et al.\ obtain sensitivity to bulk fermion parity only through their Ising-anyon
tunneling term, their fermion-tunneling term being sensitive to
$(-1)^{n_\sigma}$ alone~\cite{Klocke2022}; Wei et al.\ likewise separate
tunneling quasiparticle species~\cite{Wei2021,Wei2023}; Giuliano et al.\ state
explicitly that a two-arm interferometer accesses only the vacuum channel and
build a four-terminal device around a floating island with charging energy
$E_C$ to reach the fermionic one~\cite{Giuliano2026}; and the single-shot
readout of Ref.~\cite{MicrosoftMajorana2025} uses the quantum capacitance of a
floating island.  These examples make clear why the additional ingredients are physically
substantive rather than cosmetic.

The observable retained here is the enclosed vortex parity.  It is weaker
than a fusion measurement but more constrained than a generic spectroscopic
feature: changing localized fermion occupation does not by itself change the
vortex count, although nonequilibrium quasiparticles can broaden the resonances
and add heat-current backgrounds.  The same even-odd variable underlies the
Fabry-Perot effect in the quantum-Hall setting
~\cite{SternHalperin2006,BondersonKitaevShtengel2006}.

\subsection{Parity contrast}
\label{subsec:contrast}

\begin{figure*}[t]
  \centering
  \includegraphics[width=0.92\textwidth]{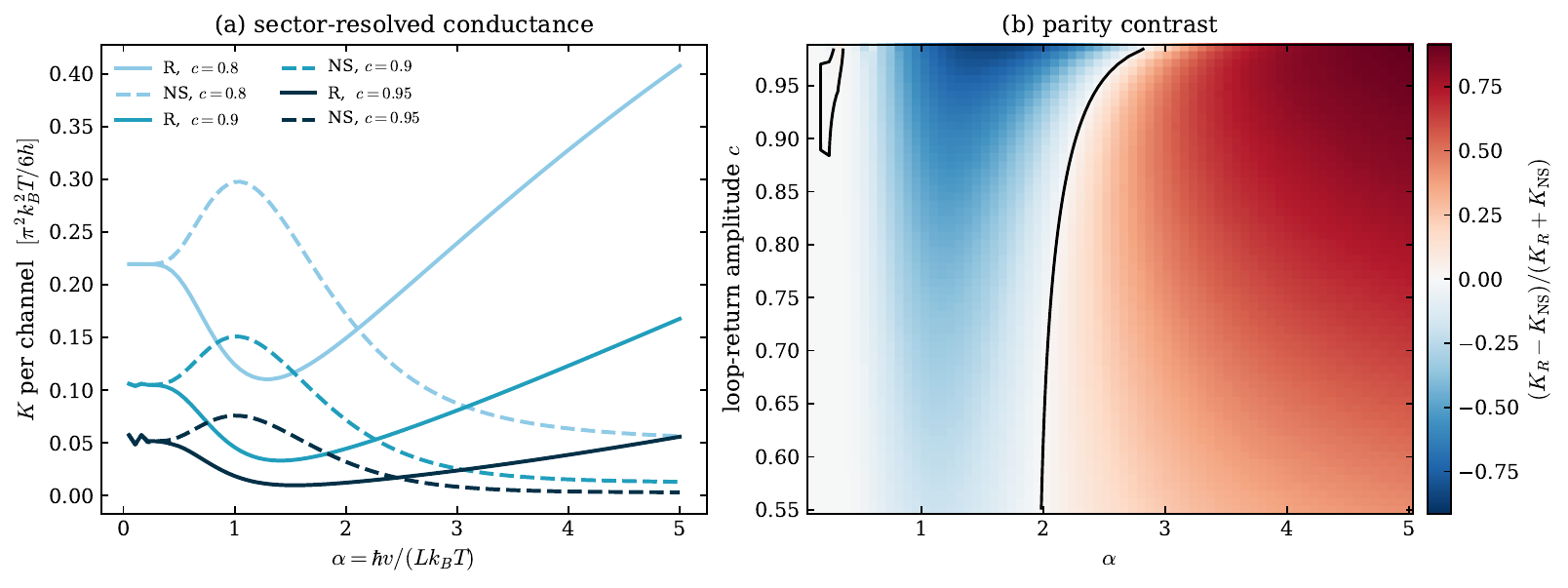}
  \caption{Parity-resolved thermal conductance of one Majorana channel, in
    units of the Majorana quantum $\pi^2\kB^2T/6h$.
    (a)~Ramond (solid) and Neveu-Schwarz (dashed) conductance versus the
    dimensionless finite-size ratio $\alpha=\hbar v/(L\kB T)$ at three
    loop-return amplitudes.  Fully transmitting contacts ($c=0$) give unity for both.
    (b)~Signed contrast $\mathcal{V}$ of Eq.~\eqref{eq:visibility} over the
    $(\alpha,c)$ plane; the black contours mark the sign reversals discussed in
    the text.  The peak magnitude is $0.91$.}
  \label{fig:contrast}
\end{figure*}

Figure~\ref{fig:contrast}(a) evaluates Eq.~\eqref{eq:K-dimensionless} for both
sectors as a function of $\alpha$ at several contact return amplitudes.  At small
$\alpha$, the thermal window samples many finite-size resonances and the two
sectors become indistinguishable.  As $\alpha$ grows the
curves separate, and the separation is large: at $\alpha\simeq1.5$ and
$c=0.95$ the Neveu--Schwarz sector conducts $5.94$ times more heat than
Ramond.

Figure~\ref{fig:contrast}(b) maps the signed contrast
\begin{equation}
  \mathcal{V} = \frac{K_R-K_{\rm NS}}{K_R+K_{\rm NS}}
\label{eq:visibility}
\end{equation}
over the $(\alpha,c)$ plane.  The dominant sign-change contour separates broad Neveu--Schwarz- and Ramond-dominated regimes.  For $\alpha\lesssim2.5$ the Neveu--Schwarz sector dominates, because its first heat-carrying level lies at half the Ramond value.  For larger $\alpha$ the ordering reverses: the Neveu--Schwarz level at $\pi\hbar v/L$ is then far outside the thermal window and exponentially suppressed, while the Ramond zero-mode resonance, though it carries no heat exactly at $\varepsilon=0$, retains weight in its wings.  The contrast exceeds $0.9$ in magnitude in the sharp-resonance corner.  Because the map extends into the very weak-contact limit $c\to1$, a second narrow sign-reversal lobe appears in the upper-left corner.  This is an isolated sharp-resonance null, not an additional phase boundary, and is avoided by a modest change of $L$, $T$, or contact coupling.  The extension to $c\simeq0.99$ is shown only to expose the sharp-resonator limit explicitly, rather than to identify an intended operating regime. Operationally this is similar to the isolated holonomy-induced nulls discussed in Appendix~\ref{app:holonomy}, although the microscopic origin is different.

Both regimes serve the experiment equally well, since what is measured is the
switch between two values under a controlled change of enclosed parity, not
the sign of the ordering.  In the independent-channel limit the sign reversal
provides an additional $\alpha$-dependent fingerprint.  Static mixing between
the two co-propagating branches can shift or even null the contrast at isolated
holonomies, as shown in Appendix~\ref{app:holonomy}; varying $L$ or $T$ then
moves the device away from such an accidental node.

\subsection{Field-sweep parity response}
\label{subsec:squarewave}

\begin{figure}[tb]
  \centering
  \includegraphics[width=0.98\columnwidth]{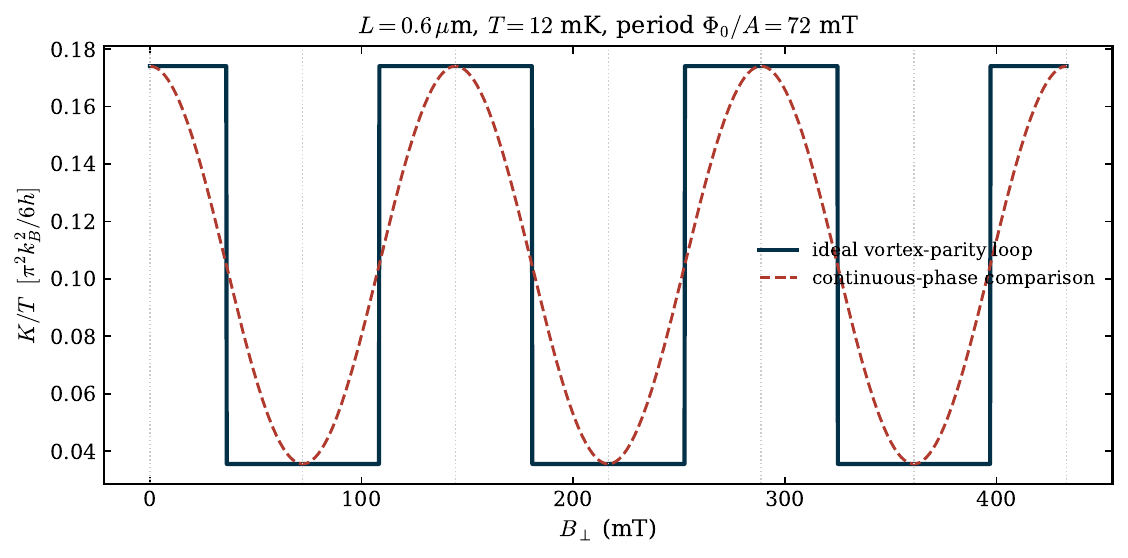}
  \caption{Idealized thermal conductance of the loop versus perpendicular
    field for the representative parameter set ($2\Delta_0/\kB T_c=20$,
    $v=1354$~m/s, $L=0.6\,\mu$m, $T=12$~mK, $c=0.93$, $|C|=1$).  Assuming one-by-one vortex entry, each parity change flips the $Z_2$ sector,
    giving a two-level response with the geometric flux scale
    $\Phi_0/A=72$~mT between $K/T=0.036$ and $0.174$ in units of the Majorana
    quantum.  A sinusoidal continuous-flux reference of the same period and
    amplitude is shown for contrast.  In experiment the vortex-entry fields
    need not be equally spaced; the signature is the conductance switch
    correlated with vortex parity.}
  \label{fig:squarewave}
\end{figure}

For a loop of area $A$, one flux quantum corresponds to the geometric scale
\begin{equation}
  \Delta B = \frac{\Phi_0}{A},\qquad \Phi_0=\frac{h}{2e}.
\label{eq:field-period}
\end{equation}
If vortices enter the enclosed region one at a time and their parity is known,
each change of $n_v$ flips $\sigma$ and hence switches the thermal conductance
between the two sector values.  In the idealized flux-counting limit this gives
the two-level waveform shown in Fig.~\ref{fig:squarewave}.

The neutral Majorana loop has no ordinary electromagnetic Aharonov--Bohm phase,
so the \emph{parity-dependent component} of the signal is discrete rather than
a continuous flux phase.  Equation~\eqref{eq:field-period} is nevertheless a
design scale, not a claim that Abrikosov vortices enter at perfectly periodic
fields.  Pinning, edge barriers and hysteresis can make the actual entry fields
irregular, while magnetic field can also change the wall profile and contacts
smoothly.  The robust test is therefore correlation of the conductance steps
with independently identified changes of enclosed vortex parity, supplemented
by the area and temperature controls below.  A recent nanoSQUID-on-tip study of
R3G/WSe$_2$ provides a concrete analysis protocol: the numerical derivative
$\mathrm{d}(\Delta B)/\mathrm{d}B_z$ separates the approximately linear
Meissner background from discrete localized vortex-entry events, whose entry
fields are not equally spaced~\cite{ZhangMeissner2026}.  The material stack is
different from the R4G/R5G devices used for the quantitative benchmarks here,
so only the identification method, not its field scales, is imported.  The
fixed-field geometric switch removes field-sweep ambiguity altogether by
changing the loop geometry at fixed field.

\subsection{Finite-size and field scales}
\label{subsec:design}

Resolving the sectors requires $\alpha\gtrsim1$, favouring a small loop.
For a field-sweep measurement it is also useful to fit several
geometric flux periods below the perpendicular critical field.  Imposing the
conservative guideline $\Delta B\lesssim B_c^\perp/3$ favours a larger area.
For a circular loop, which maximizes area at fixed perimeter, $A=L^2/4\pi$ and
this field-sweep guideline reads
\begin{equation}
  L \;\gtrsim\; \sqrt{\frac{12\pi\Phi_0}{B_c^\perp}} .
\label{eq:L-min}
\end{equation}

\begin{figure}[tb]
  \centering
  \includegraphics[width=0.98\columnwidth]{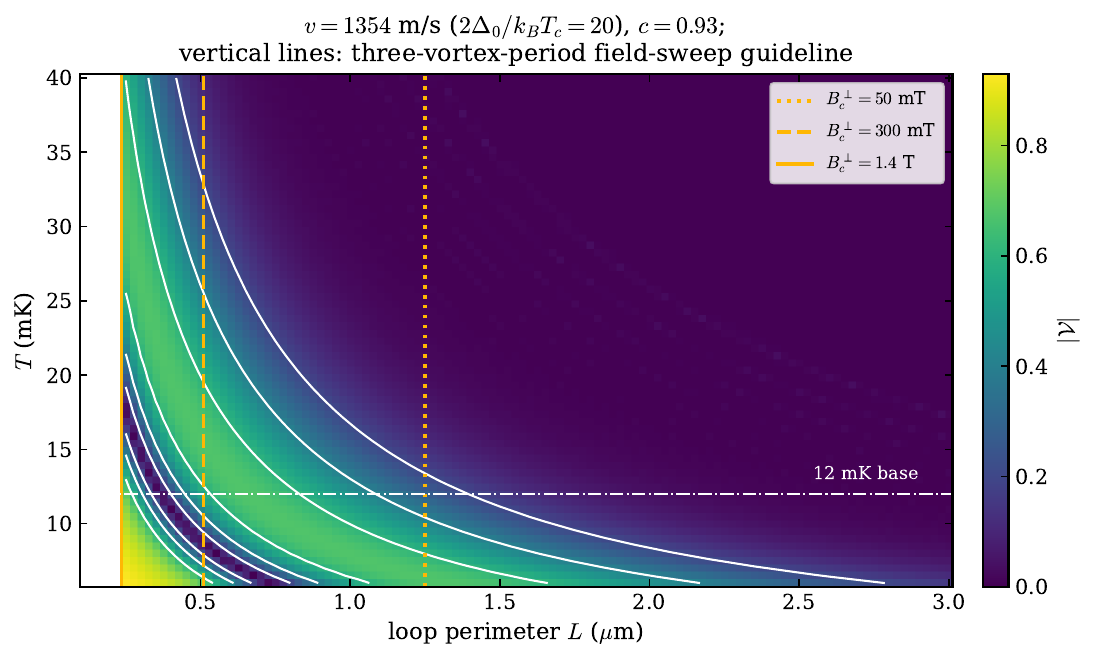}
  \caption{Finite-size and field map.  Colour is the magnitude of the parity contrast
    $|\mathcal{V}|$ at $c=0.93$ for $v=1354$~m/s
    ($2\Delta_0/\kB T_c=20$).  White contours are $|\mathcal{V}|=0.2$, $0.4$
    and $0.6$.  Vertical lines give the minimum perimeter of
    Eq.~\eqref{eq:L-min} for three perpendicular critical fields; the
    field-sweep-compatible region lies to their right and inside the
    high-contrast region.
    The horizontal line marks the $12$~mK base temperature of existing
    graphene noise thermometry.}
  \label{fig:design}
\end{figure}

Figure~\ref{fig:design} maps the contrast over loop perimeter and temperature
at fixed material parameters, with the field-sweep scale
Eq.~\eqref{eq:L-min} overlaid for three values of $B_c^\perp$.  At the
$12$~mK base temperature the independent-channel contrast satisfies
$|\mathcal V|>0.3$ over two intervals in the plotted range,
$0.25\lesssim L\lesssim0.32\,\mu$m and
$0.42\lesssim L\lesssim1.22\,\mu$m.  They are separated by the narrow
accidental contrast node near $L\simeq0.365\,\mu$m discussed above.  The geometric
three-period criterion requires $L\geq1.25$, $0.51$ and $0.24\,\mu$m for
$B_c^\perp=50$~mT, $300$~mT and $1.4$~T, respectively.  The last value is the
critical perpendicular field reported in the chiral-superconducting
rhombohedral-graphene devices of Ref.~\cite{Han2025}, so the representative
$0.6\,\mu$m loop lies comfortably inside the field budget.  The fixed-field geometric switch is not subject to this field-sweep constraint.

Table~\ref{tab:design} collects a representative parameter set.  The absolute conductances are approximately $2\times10^{-16}$ and
$10^{-15}$~W/K in the two sectors at $12$~mK.  Graphene Johnson-noise
thermometry has resolved quantized thermal conductance to about $5\%$ accuracy
at millikelvin temperatures~\cite{Srivastav2019}; a complementary non-local
graphene noise thermometer reached a precision of about $1\%$ of a thermal
conductance quantum at $5$~K~\cite{Waissman2021}.

\begin{table}[tb]
  \centering
  \caption{Representative parameter set and derived scales.  We use
    $T_c=300$~mK from the chiral-superconducting R4G/R5G devices
    \cite{Han2025} and the experimentally quoted $k_F=0.29$~nm$^{-1}$ from
    the degeneracy-two R8G Landau fan~\cite{Okounkova2026} as a conservative
    rhombohedral-graphene momentum scale.  The gap ratio
    $2\Delta_0/\kB T_c=20$ is the illustrative benchmark of
    Ref.~\cite{Ghosh2026THT}, not a spectroscopic determination.}
  \label{tab:design}
  \begin{ruledtabular}
  \begin{tabular}{ll}
    quantity & value \\
    \hline
    pairing gap, $\Delta_0$ & $0.2585$~meV \\
    Majorana edge velocity, $v=\Delta_0/\hbar k_F$ & $1354$~m/s \\
    thermal wavelength at $12$~mK, $\lambda_T$ & $5.4\,\mu$m \\
    loop perimeter, $L$ & $0.60\,\mu$m \\
    enclosed area, $A=L^2/4\pi$ & $0.029\,\mu$m$^2$ \\
    finite-size ratio, $\alpha$ & $1.44$ \\
    geometric flux scale, $\Phi_0/A$ & $72$~mT \\
    two-branch loop NS conductance, $K/T$ & $0.174\,\pi^2\kB^2/6h$ \\
    two-branch loop R conductance, $K/T$ & $0.036\,\pi^2\kB^2/6h$ \\
    ratio, $K_{\rm NS}/K_R$ & $4.9$ \\
    open-contact plateau, $K_{\rm DW}/T$ & $9.46\times10^{-13}$~W/K$^2$ \\
  \end{tabular}
  \end{ruledtabular}
\end{table}

\subsection{Thermal backgrounds}
\label{subsec:phonons}

Phonon conduction is the central obstacle for thermal probes of Kitaev
magnets, where it dominates the edge signal and forces a dedicated device
architecture of mesoscopic and macroscopic regions to manage
it~\cite{Klocke2022}.  The floating-contact geometry changes that background problem substantially.

The floating-contact method does not infer the signal from a macroscopic
temperature gradient.  It measures the heat balance of a small ohmic contact
whose electron temperature is read by Johnson-noise thermometry.  In existing
graphene implementations the dominant electron-phonon relaxation of that
contact is calibrated from its distinct power law, while additional heat leaks
enter as backgrounds to be characterized independently.  Srivastav et al.\
extract quantized edge contributions in hBN-encapsulated graphene at
$\sim12$~mK in precisely this measurement architecture
~\cite{Srivastav2019,Srivastav2022}, and Banerjee et al.\ resolved individual
thermal quanta in the same way~\cite{Banerjee2017,Banerjee2018}.  Two features provide additional discrimination.  The signal is not a small
correction to a large background but a parity-correlated switch of the loop
conductance, and it changes when a matched contour is moved across a fixed
vortex while the contact and field settings are held fixed.  Phonon and static
contact backgrounds are not tied to that enclosed-parity change.  Differencing
the two matched geometries therefore suppresses such backgrounds without
assigning them a topological origin.

\subsection{Rhombohedral-graphene scale comparison}
\label{subsec:material}

Within the weak-pairing benchmark, the key material scale controlling the
finite-size resonance is the chiral Majorana edge velocity
$v=\Delta_0/\hbar k_F$.  The sense of this dependence is opposite
to intuition drawn from Fermi velocities: a \emph{small} Fermi wavevector
\emph{enhances} $v$.  Rhombohedral graphene, using the representative
$k_F=0.29$~nm$^{-1}$ from the degeneracy-two Landau fan
~\cite{Okounkova2026}, has the smallest momentum scale among the representative
platforms in Table~\ref{tab:materials} and therefore a comparatively large
$v$ despite its small gap.

Table~\ref{tab:materials} gives an illustrative scale comparison.  As a conservative
scale-separation diagnostic, we report $L(\alpha=1.5)/(10\xi)$; values above
unity place the loop at least an order of magnitude beyond the quoted bulk
coherence scale, but this ratio is not a universal domain-wall-width criterion.
The same perimeter must also lie at a scale on which chiral-domain textures can
be imaged and reconfigured.  For the representative parameters in Table~\ref{tab:materials}, the resolving
loops of Sr$_2$RuO$_4$ and UTe$_2$ fall below this conservative separation
scale, while at $L=1\,\mu$m both have $\alpha\lesssim0.15$.  The Fe(Te,Se)
surface benchmark clears the scale but offers no demonstrated rewritable
chiral domains.  For RHG, the $1.4$~T perpendicular critical field reported in
Ref.~\cite{Han2025} corresponds to the orbital scale
$\xi_B=\sqrt{\Phi_0/(2\pi B_c^\perp)}\simeq15.3$~nm, giving a margin of about
$3.75$ at $L(\alpha=1.5)\simeq575$~nm.  This perimeter lies in the mesoscopic
range addressed by nanoscale imaging and reconfiguration of chiral domains
~\cite{Dutta2026,Hua2026}.

\begin{table}[tb]
  \centering
  \caption{Representative superconducting scale comparisons relevant to
    thermal Majorana interferometry, at $T=12$~mK.  Here $v=\Delta_0/\hbar k_F$;
    $L(\alpha{=}1.5)$ is the loop perimeter needed to resolve the two sectors;
    and the margin is $L(\alpha{=}1.5)/10\xi$, used here as a
    conservative scale-separation diagnostic rather than a universal wall-width
    criterion.  For RHG we use the orbital scale inferred from
    $B_c^\perp=1.4$~T~\cite{Han2025}; the other coherence lengths are
    representative literature values~\cite{Mackenzie2003,Ran2019,Wang2018FeTeSe}.
    These rows are scale comparisons only and do not imply established chiral
    topological superconductivity or rewritable domains in the comparison materials.}
  \label{tab:materials}
  \begin{ruledtabular}
  \begin{tabular}{lccccc}
    & $\Delta_0$ & $k_F$ & $v$ & $L(\alpha{=}1.5)$ & margin \\
    & (meV) & (nm$^{-1}$) & (m/s) & (nm) & \\
    \hline
    Sr$_2$RuO$_4$      & 0.30  & 6.0  & \phantom{0}76 & \phantom{0}32 & 0.05 \\
    UTe$_2$            & 1.0\phantom{00} & 7.0 & 217 & \phantom{0}92 & 0.77 \\
    Fe(Te,Se) surface  & 0.20  & 1.5  & 203 & \phantom{0}86 & 3.4\phantom{0} \\
    RHG benchmark        & 0.259 & 0.29 & 1354 & 575 & 3.75 \\
  \end{tabular}
  \end{ruledtabular}
\end{table}

\subsection{Edge-vortex hybridization: an intrinsic false positive}
\label{subsec:artefact}

The no-go of Sec.~\ref{subsec:nogo} says the fusion channel is invisible.  It
does not say that everything visible is the enclosed vortex parity, and the
difference matters, because there is a systematic that mimics the signal.

A vortex core lying a distance $d$ from the wall hybridizes with the chiral
branch through a matrix element $A_{\eta\gamma}\neq0$.  We parameterize the
resulting resonance by a width whose distance dependence is estimated as
\begin{equation}
  \Gamma \sim \Gamma_0 e^{-2d/\xi},\qquad \Gamma_0\sim\frac{\hbar v}{L}.
\label{eq:hyb-width}
\end{equation}
Chirality forbids reflection, so the transmission amplitude past such a
side-coupled zero mode is unimodular,
\begin{equation}
  s(\varepsilon)
  = \frac{\varepsilon-i\Gamma/2}{\varepsilon+i\Gamma/2},
  \qquad
  \arg s = -2\arctan\!\frac{\Gamma}{2\varepsilon},
\label{eq:mzm-phase}
\end{equation}
which is a $\pi$ phase shift at $\varepsilon=0$ decaying to zero for
$|\varepsilon|\gg\Gamma$ (Fig.~\ref{fig:artefact}a).  Each such mode multiplies
the round-trip amplitude in Eq.~\eqref{eq:transmission} by $s(\varepsilon)$.

This yields a direct false-positive mechanism.  A vortex sitting just
\emph{outside} the contour, but within $\xi$ of it, contributes a low-energy
$\pi$ shift indistinguishable from that of an \emph{enclosed} vortex.
Figure~\ref{fig:artefact}b quantifies it: the spurious contrast grows from
$0.9\%$ of the true value at $\Gamma/\kB T=0.05$ to $89.6\%$ at
$\Gamma/\kB T=20$.  Since Eq.~\eqref{eq:hyb-width} gives
$\Gamma/(\hbar v/L)=0.37$, $0.14$ and $0.02$ at $d/\xi=0.5$, $1$ and $2$, a
standoff of two coherence lengths suppresses the artefact to the percent level
for this prefactor estimate, whereas a vortex pinned against the wall need not
be perturbative.

Two complementary discriminators separate the two mechanisms.

\emph{Temperature dependence.}  The true switch multiplies the entire round
trip by $\sigma$ and is energy independent.  At fixed, independently calibrated
contact coupling, $\mathcal{V}_{\rm true}$ depends only on $\alpha$; for the
representative $c=0.93$ used in Fig.~\ref{fig:artefact} it changes sign at
$\alpha=2.36$.  The artefact is energy dependent and requires a second
parameter $\gamma=\Gamma L/\hbar v$; its curves neither collapse onto the
reference curve nor share its zero crossing (Fig.~\ref{fig:artefact}c).  A temperature sweep therefore provides a constrained fingerprint once the
contact coupling is calibrated independently.

\emph{Geometry.}  The fixed-field geometric switch moves the contour so that
a vortex passes from inside to outside at matched standoff.  The true parity
switch flips, whereas the leading hybridization scale, controlled
exponentially by $d$, can be held approximately fixed.  This provides the
strongest discriminator and is available only in a rewritable device.

\subsection{Full counting statistics: a mean--noise falsification test}
\label{subsec:fcs}

The hybridization artefact of Sec.~\ref{subsec:artefact} is spectrally different
from a genuine change of spin structure, even when a single integrated
observable happens to coincide.  Heat full counting statistics makes that
difference explicit.  The static, elastic limit of the driven cumulant
generating function of Ref.~\cite{Simons2020}, which in turn extends the
scattering formalism of Muzykantskii and Khmelnitskii~\cite{Muzykantskii1994}
from charge to energy transfer, reduces for a frozen transmission to the
Levitov--Lesovik form~\cite{LevitovLesovik1993,Klich2003}.  Restricting the
energy integral to positive energies, which is the statement that a Majorana
branch carries half the independent excitations of a Dirac channel, the
cumulant generating function for measurement time $\tau_m$ and reservoir
temperatures $T_L,T_R$ is
\begin{equation}
\begin{split}
 \frac{\mathcal F_\sigma(\lambda)}{\tau_m}
 &=\frac{1}{h}\int_0^\infty \!\dd E\,
 \ln\!\Bigl\{1+\mathcal T_\sigma(E)\Bigl[
 A_E\left(e^{i\lambda E}-1\right)\\
 &\hspace{7.6em}
 +B_E\left(e^{-i\lambda E}-1\right)\Bigr]\Bigr\}.
\end{split}
\label{eq:fcs-cgf}
\end{equation}
where $A_E=f_L(1-f_R)$, $B_E=f_R(1-f_L)$ and
$\mathcal T_\sigma(E)$ is Eq.~\eqref{eq:transmission}, including the
energy-dependent phase of Eq.~\eqref{eq:mzm-phase} when a nearby vortex is
present.  The first two cumulant rates are therefore
\begin{align}
 J_\sigma & = \frac{1}{h}\int_0^\infty \!\dd E\,E\,
 \mathcal T_\sigma(E)\,[f_L-f_R], \label{eq:fcs-mean}\\
 S_{Q,\sigma} & = \frac{1}{h}\int_0^\infty \!\dd E\,E^2
 \left\{\mathcal T_\sigma[A_E+B_E]
 -\mathcal T_\sigma^2[f_L-f_R]^2\right\}. \label{eq:fcs-noise}
\end{align}
Here $S_{Q,\sigma}$ denotes the second cumulant of transferred energy per unit
time.  With this convention the transparent limit $\mathcal T_\sigma\equiv1$ returns
$K=\pi^2\kB^2T/6h$, one Majorana thermal quantum, so
Eqs.~\eqref{eq:fcs-mean} and~\eqref{eq:fcs-noise} are normalized consistently
with the rest of the paper.  At equilibrium, Eq.~\eqref{eq:fcs-noise} reduces
exactly to the thermal fluctuation--dissipation relation
\begin{equation}
 S_{Q,\sigma}^{\rm eq}=2\kB T^2 K_\sigma .
\label{eq:fcs-fdt}
\end{equation}
Consequently equilibrium noise contains no information independent of the
linear conductance.  The additional discriminator appears only away from
linear response, where the $\mathcal T_\sigma^2$ partition term probes the
energy dependence of the resonance.  Equation~\eqref{eq:fcs-cgf} also obeys
the Gallavotti--Cohen symmetry
$\mathcal F_\sigma(\lambda)=\mathcal F_\sigma[-\lambda+i(\beta_R-\beta_L)]$;
at $T_L=T_R$ it is even in $\lambda$, as required for an unbiased static
scatterer.

\begin{figure*}[t]
  \centering
  \includegraphics[width=0.88\textwidth]{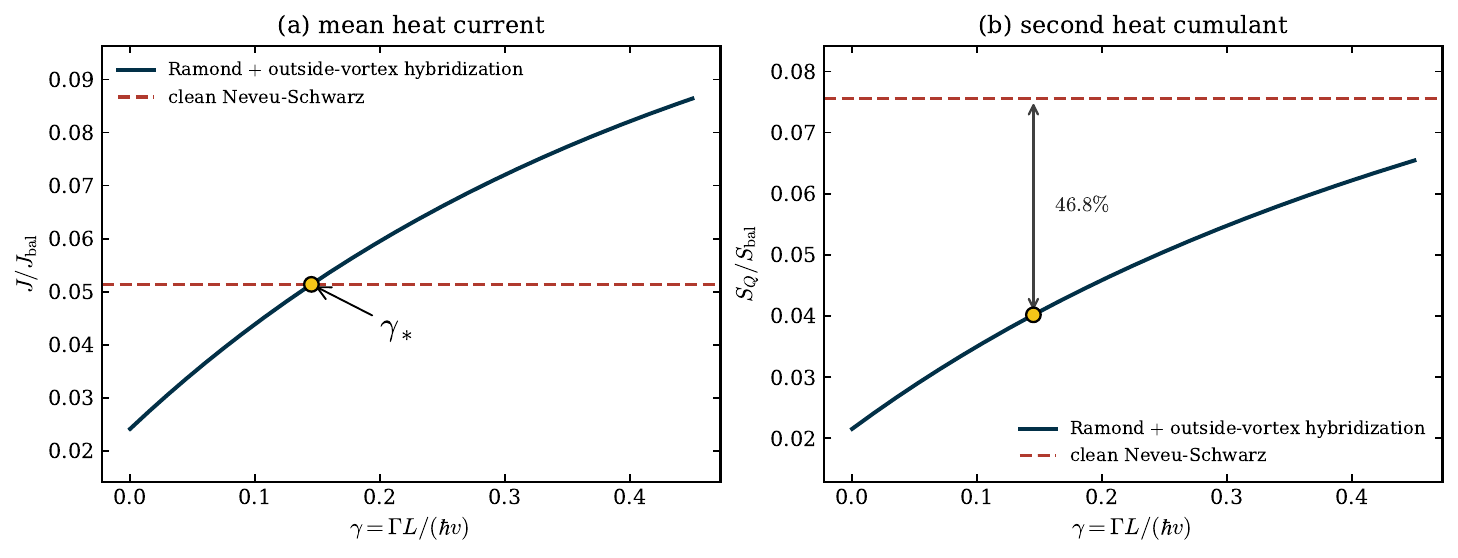}
  \caption{Full-counting-statistics discriminator for an intrinsic
  false positive.  Parameters are $\alpha=3$, $c_1=c_2=0.93$ and
  $T_L/T_R=2$.  Solid curves show a Ramond loop with one side-coupled vortex
  outside the contour as a function of
  $\gamma=\Gamma L/(\hbar v)$; dashed lines show a clean Neveu--Schwarz loop.
  (a)~The mean heat currents cross at $\gamma_*=0.14510$ (marker), so the two
  physically distinct spectra are indistinguishable by the average current
  alone.  (b)~At precisely the same point the normalized second heat cumulants
  differ by $46.8\%$.  Both normalizations are the transparent one-Majorana
  values for the same pair of reservoir temperatures.}
  \label{fig:fcs}
\end{figure*}

Figure~\ref{fig:fcs} isolates a mean-degenerate false positive.  We choose
$\alpha=3$, $c_1=c_2=0.93$ and $T_L=2T_R$.  For a clean Neveu--Schwarz loop,
the normalized mean heat current is $J/J_{\rm bal}=0.051386$, where
$J_{\rm bal}=\pi^2\kB^2(T_L^2-T_R^2)/(12h)$ is the transparent-Majorana value.
A Ramond loop with one vortex outside the contour can be tuned to the
\emph{same} mean current at
\begin{equation}
 \gamma_*\equiv\frac{\Gamma L}{\hbar v}=0.14510 .
\label{eq:fcs-degenerate}
\end{equation}
With the prefactor estimate of Eq.~\eqref{eq:hyb-width}, this corresponds to
$d/\xi\simeq0.97$, so the example lies in the intrinsic false-positive regime
rather than an artificially large coupling.  At the matched-mean point the
second cumulant remains sharply different:
$S_Q/S_{\rm bal}=0.04019$ for the hybridized Ramond loop versus $0.07556$ for
the clean Neveu--Schwarz loop, a $46.8\%$ separation.  Here
$S_{\rm bal}=\pi^2\kB^3(T_L^3+T_R^3)/(6h)$.  Thus a finite-bias noise
measurement can falsify a mean-degenerate hybridization mimic without invoking
fusion-channel sensitivity or any time-dependent braiding operation.

\subsection{Partial coherence}
\label{subsec:coherence}

The field-driven and fixed-field measurements require phase coherence over the loop.  Rather than assume
it, we quantify the cost of losing it.  Separating diagonal from off-diagonal
path pairs in
$|1-ue^{i\theta}|^{-2}=\sum_{m,n}u^{m+n}e^{i(m-n)\theta}$, with
$u=c_1c_2\sigma$, and damping the interference terms phenomenologically by
$\lambda^{|m-n|}$, in the spirit of a B\"uttiker dephasing
probe~\cite{Buttiker1986}, gives a series that sums exactly to
\begin{equation}
  T=\frac{T_1T_2\left(1-\lambda^2u^2\right)}
       {\left(1-u^2\right)\left(1-2\lambda u\cos\theta+\lambda^2u^2\right)},
\label{eq:dephased-T}
\end{equation}
with $\lambda=e^{-L/\ell_\phi}$ the amplitude coherence per round trip.
Equation~\eqref{eq:dephased-T} is exact within this phenomenological damping
prescription, reduces to Eq.~\eqref{eq:transmission} at $\lambda=1$, gives the classical
parity-blind ladder at $\lambda=0$, and returns unity for transparent contacts
at every $\lambda$.

\begin{figure*}[t]
  \centering
  \includegraphics[width=0.95\textwidth]{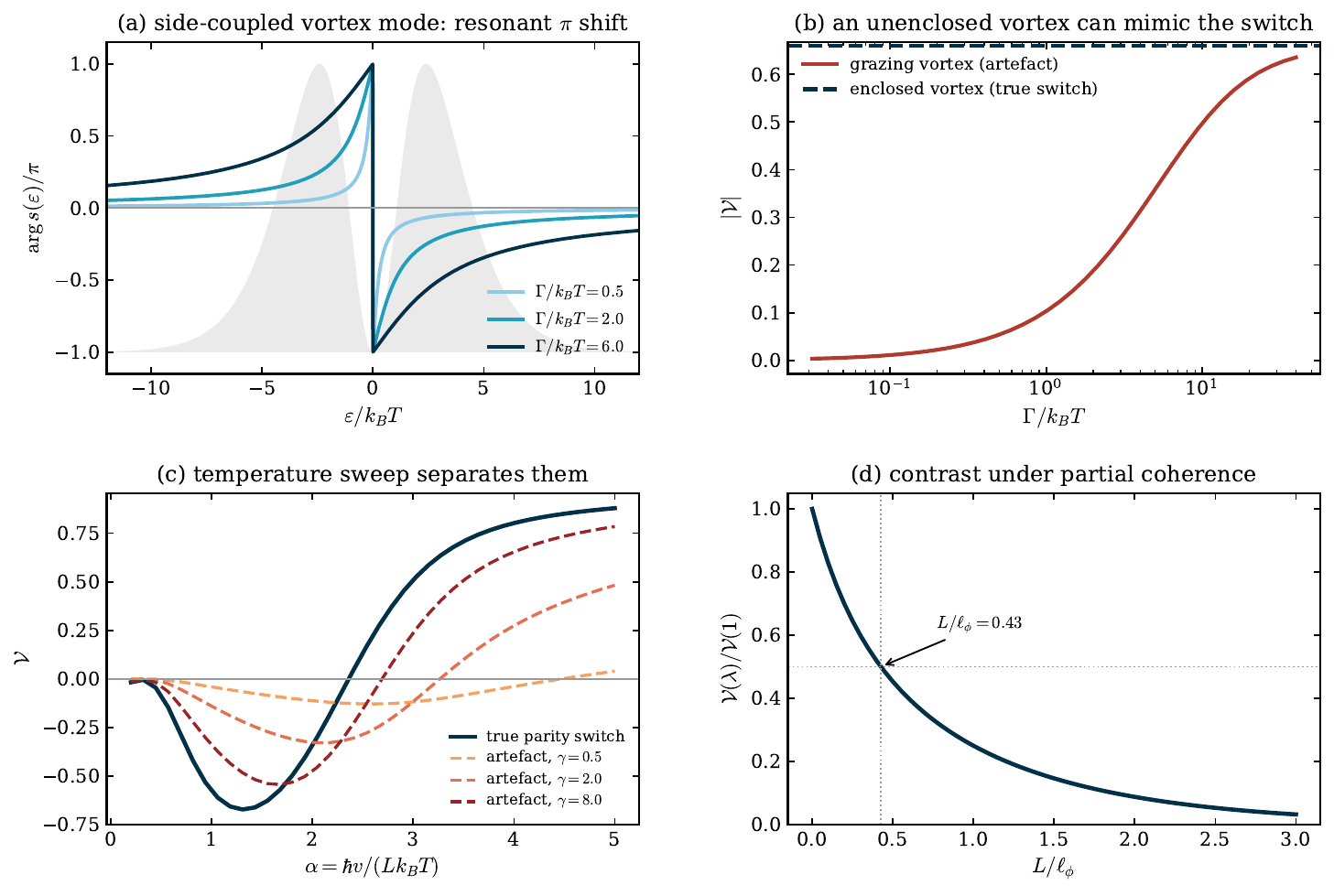}
  \caption{Systematics of the parity readout.
    (a)~Transmission phase past a side-coupled vortex Majorana mode,
    Eq.~\eqref{eq:mzm-phase}: a $\pi$ shift at $\varepsilon=0$ decaying for
    $|\varepsilon|\gg\Gamma$.  Shaded, the thermal weight.
    (b)~A vortex outside the contour but hybridized with it produces a
    spurious contrast that approaches the true parity switch (dashed) as
    $\Gamma/\kB T$ grows.
    (c)~Discriminator.  At the fixed contact setting $c_1=c_2=0.93$,
    the true switch depends only on $\alpha$ and crosses zero at
    $\alpha=2.36$; the artefact requires the additional parameter
    $\gamma=\Gamma L/\hbar v$ and does not collapse onto the same curve.
    (d)~Contrast under partial coherence from the closed-form
    phenomenological dephasing model of Eq.~\eqref{eq:dephased-T}; it halves
    at $L/\ell_\phi=0.43$.
    All panels use $c_1=c_2=0.93$ and, where fixed, $\alpha=1.44$.}
  \label{fig:artefact}
\end{figure*}

At the representative point $\alpha=1.44$ and $c_1=c_2=0.93$,
Fig.~\ref{fig:artefact}d shows that the contrast falls to half its coherent
value at
\begin{equation}
  L/\ell_\phi = 0.43,
\label{eq:coherence-requirement}
\end{equation}
corresponding to $\ell_\phi\simeq2.3L$ for half contrast at this parameter
set; a quarter of the coherent contrast remains when $\ell_\phi=L$.  The
degradation is gradual, but $\ell_\phi$ has not yet been measured for an RHG
chiral Majorana wall.  In a clean chiral-Majorana edge theory, charge
conjugation forbids the ordinary electron--phonon coupling and the lowest local
self-interaction is highly irrelevant~\cite{Lian2018,FuKane2009}, so intrinsic
inelastic scattering is expected to be weak at low temperature.  This makes
Eq.~\eqref{eq:coherence-requirement} plausible as an intrinsic-inelastic
constraint, but not as a bound on the total RHG coherence length: wall
roughness, interbranch mixing and coupling to additional low-energy modes may
still dominate.  The long thermal wavelength also suggests reduced
sensitivity to very short-scale wall structure.  These considerations do not
replace a direct measurement of $\ell_\phi$.

%%==========================================================================
\section{Experimental consequences and controls}
\label{sec:protocol}
%%==========================================================================

The theory yields three thermal tests on the same device and one optional
electrical cross-check requiring an additional charged edge channel.

\subsection{Open-contact plateau calibration}
\label{subsec:calib}

With both point contacts fully open, the low-temperature gapped regime of
Eq.~\eqref{eq:plateau-limit} gives
$K_{\rm DW}/T=|C|\,\pi^2\kB^2/3h$, independent of loop size and insensitive to
weak disorder that does not change the chiral mode count, provided
Eq.~\eqref{eq:gapped} holds and the contacts are effectively transparent.  Erasing
the domain removes the channel entirely.  This calibration fixes $|C|$,
verifies the thermometry against a known quantum, and establishes that the
written wall carries the predicted number of modes before the resonant regime
is examined.  It is also the direct test of the domain-wall
prediction of Ref.~\cite{Ghosh2026THT}.

\subsection{Field-driven vortex-parity steps}
\label{subsec:field}

With the contacts tuned to $c\simeq0.9$--$0.95$, a sweep of $B_\perp$ gives
the two-level response of Fig.~\ref{fig:squarewave} in the idealized
one-by-one entry limit, although the vortex-entry fields need not form a
perfect arithmetic sequence.  The essential measurement is a
thermal-conductance switch each time the \emph{enclosed vortex parity} changes.
Direct nanoscale magnetometry, where available, can identify the vortex
configuration; otherwise repeated cool-downs and the $A$ dependence of the
mean entry scale provide controls.

An important property of this measurement is that it accesses the parity of the
\emph{vortex number}, not the occupation parity of a localized fermion.  A
stray quasiparticle can change fermion occupation without changing how many
vortices are enclosed.  Slow thermal integration therefore does not by itself
average away the geometric vortex-parity label, although nonequilibrium
quasiparticles can still broaden resonances and add background heat transport.
Three quantitative checks accompany the measurement.  First, resizing the
loop must change the characteristic flux scale as $1/A$.  Second, opening the
contacts must collapse the parity contrast toward the plateau of the open-contact calibration.
Third, the contrast must evolve with $L/T$ according to the resonance theory,
including any static interbranch-holonomy offset calibrated as in
Appendix~\ref{app:holonomy}.

\subsection{Fixed-field geometric parity switch}
\label{subsec:geometry}

A fixed-field geometry removes the principal field-sweep ambiguity.  At fixed
$B_\perp$, a
pinned vortex at a known position defines two nearly equal-perimeter contours,
one excluding and one enclosing the same vortex.  Reconfiguration of the
closed chiral-domain wall between these contours leaves the bulk state,
magnetic field and contact settings unchanged while only
$n_v\bmod2$ changes.  Equations
\eqref{eq:z2} and \eqref{eq:transmission} then predict a switch between the
Neveu--Schwarz and Ramond thermal conductances.  A matched contour deformation that does \emph{not} cross a vortex provides
the corresponding null control.

This fixed-field geometry removes the strongest field-sweep ambiguity: vortex
pinning and smooth magnetoresponse are held fixed while the same imaged vortex
is placed inside or outside a rewritable contour.  Geometry-dependent
backgrounds are constrained separately by the matched null deformation.  The required capability is more specific than what is
currently demonstrated.  Ref.~\cite{Dutta2026} establishes deterministic
ultra-low-current reconfiguration of chiral domains, but not arbitrary
micron-scale contour writing around an individual Abrikosov vortex.
Independently, chip-integrated control loops beneath NbSe$_2$ have positioned
individual Abrikosov vortices to better than $100$~nm, pushed or pulled them
over distances up to $3~\mu$m, and implemented shuttle and winding sequences
under SQUID-on-Tip imaging~\cite{Keren2023}.  This does not demonstrate the
operation in rhombohedral graphene or its integration with a rewritable chiral
wall, but it narrows the missing capability to combining established
single-vortex control with local domain-wall writing.  Local gates or current
lines able to pin two matched contours remain an explicit fabrication
requirement rather than an assumed experimental fact.

\subsection{Optional charge-based cross-check}
\label{subsec:charge}

The charge interferometers of Fu--Kane and Akhmerov--Nilsson--Beenakker provide
an independent way to read vortex parity when a \emph{genuine charged chiral
Dirac channel} is available~\cite{FuKane2009,Akhmerov2009,Alicea2012}.  In the
ideal balanced converter geometry their zero-bias result is
\begin{equation}
  G(0) = \begin{cases}
    0, & n_v \ \text{even},\\[2pt]
    2e^2/h, & n_v \ \text{odd},
  \end{cases}
\label{eq:charge-readout}
\end{equation}
up to the convention for which drain is labelled electron-like.  This result is included only as an optional cross-check.

A wall between superconducting regions with $\CBdG=+1$ and $-1$ carries two
neutral Majorana branches.  Combining those operators into a complex fermion
does \emph{not} by itself create a charge-$e$, $U(1)$-conserving Dirac lead.
Implementing Eq.~\eqref{eq:charge-readout} in rhombohedral graphene would
therefore require a separately established charged QAH or normal chiral edge
that can be converted into the Majorana branches.  Quantum anomalous Hall and
superconducting phases are both known in the rhombohedral-graphene family
~\cite{Choi2025}, but their local coexistence in the required converter
geometry has not been demonstrated.  The charge-based cross-check is consequently not part of the core thermal
analysis.

If such a charged converter is realized, simultaneous charge and heat readout
would be valuable because the systematic errors are different.  The thermal
resonator probes a shifted energy spectrum, while the charge interferometer
probes electron--hole conversion.  Agreement of their parity switches at the
same identified vortex crossing would then be a stringent cross-check rather
than a prerequisite for the interferometer.

\subsection{Falsification criteria}
\label{subsec:falsify}

A topological interpretation is not supported if the parity-correlated component
cannot be separated from smooth backgrounds or if the response fails to track
controlled changes of enclosed vortex parity.  A purely sinusoidal
response that persists when the enclosed vortex configuration is held fixed
would indicate an ordinary continuous interferometric phase rather than the
neutral $Z_2$ boundary-condition switch.  Likewise, the characteristic field
scale should evolve with area even though individual vortex-entry fields may
be irregular.  A plateau in the open-contact calibration at other than an
integer multiple of $\pi^2\kB^2T/3h$, or a reproducible electronic residual
linear-in-$T$ longitudinal thermal conductance, would invalidate the quantized
fully gapped interpretation until contact, background and gappedness checks are
satisfied; a Bogoliubov Fermi surface is one possible microscopic origin.  In that regime the quantized-plateau premise of
Ref.~\cite{Ghosh2026THT} no longer applies, so the appropriate response is to
identify a different fully gapped gate setting rather than reinterpret a
nonquantized value as a Chern number.  A useful additional control follows the
logic of the trivial-phase check in Ref.~\cite{MicrosoftMajorana2025}: tune the
two sides of the written contour into the same Chern phase so that
$\Delta C=0$.  The domain-wall channel then disappears, and with it any
parity-correlated resonator response, while the contact layout and geometric
area remain essentially unchanged.  An even-$|C|$ phase is not a strict null control.  A physical $h/2e$ vortex can
still alter the fermionic boundary condition, but it carries no topologically
protected unpaired core Majorana mode and generic interbranch mixing can remove
a zero-energy resonance.  The clean null control is instead $\Delta C=0$,
where the domain-wall channel itself disappears.

%%==========================================================================
\section{Discussion}
\label{sec:discussion}
%%==========================================================================

The construction rests on a single observation: for odd $\CBdG$, changing the
parity of the number of enclosed unit vortices changes the spin structure and
therefore shifts the resonance spectrum of a closed chiral Majorana edge by
half a level spacing.  Heat transport reads that spectrum.  What makes this
potentially practical in rhombohedral graphene is the small Majorana edge
velocity $\Delta_0/\hbar k_F\sim10^3$~m/s, which places the finite-size scale in
the millikelvin range for submicron-to-micron contours.

\paragraph{Thermal interferometry of neutral anyons.}
The closest theoretical precedent for heat-based neutral-anyon interferometry
is found in spin liquids rather than in nanowire parity metrology.  Wei,
Mitrovi\'c and Feldman established that Fabry-P\'erot and Mach-Zehnder heat
interferometers distinguish topological orders through the statistics of the
tunneling quasiparticle, and that the two geometries behave qualitatively
differently~\cite{Wei2021,Wei2023}.  Klocke, Moore, Alicea and Hal\'asz then
designed mixed mesoscopic-macroscopic devices for Kitaev magnets that extract
the quantized edge conductance, identify Ising anyons from the temperature
dependence, and detect single bulk anyons through a two-pinch braiding
term~\cite{Klocke2022}.  Their fermion-tunneling correction carries the factor
$(-1)^{n_\sigma}$, the same enclosed-vortex parity measured here in a
different material.

The present geometry differs in several experimentally relevant respects.  Their braiding term is a perturbative
$\mathcal{O}(t_Lt_R)$ correction to $\kappa_e$ and is exponentially suppressed
once the pinch separation exceeds the thermal length; the present signal is
non-perturbative in the contacts, resummed to all orders in round trips, and is
a switch of the entire loop conductance between two values.  Their architecture
exists to manage phonons, which in $\alpha$-RuCl$_3$ dominate the edge
signal~\cite{Klocke2022}; the floating-contact geometry here removes that
requirement (Sec.~\ref{subsec:phonons}).  Their geometry is lithographic, so
the enclosed anyon content cannot be changed at fixed field, whereas
the fixed-field geometric switch toggles it by rewriting the contour, which also provides a strong
control against the hybridization artefact of Sec.~\ref{subsec:artefact}.  Finally, the bulk invariant is calibrated in situ by the open-contact plateau
on the same wall.  Relative to the spin-liquid interferometers of
Ref.~\cite{Klocke2022}, the distinctive ingredients are the rewritable contour
and the non-perturbative resonator regime.  Relative to the X-shaped device of
Ref.~\cite{Giuliano2026}, no floating island, edge-vortex tunneling or charge
measurement is required, with the deliberately narrower information content
made precise in Sec.~\ref{subsec:nogo}: vortex parity rather than the fusion
channel.

An instructive comparison is the nanowire parity-readout experiment of
Ref.~\cite{MicrosoftMajorana2025}, which achieves microsecond single-shot
fermion-parity readout with a reported $1\%$ assignment error and separates
that measurement from a topological-gap protocol.  The present thermal geometry
does not compete on speed or single-shot fidelity.  Instead, the open-contact
plateau fixes the number of chiral domain-wall modes through a quantized thermal
conductance, after which the parity-sensitive measurements test the
vortex-induced boundary-condition switch in the same two-dimensional chiral
phase.  The distinguishing feature is thus a direct bulk--edge thermal
calibration tied to the same domain-wall object used for the parity test.

Three further features follow from the material rather than the readout
architecture.  There is no proximity interface, no required applied in-plane
field and no spin-orbit engineering, because the chiral superconductivity is
intrinsic~\cite{Han2025}.  This removes ingredients required by the nanowire
implementation of Ref.~\cite{MicrosoftMajorana2025}, although it does not by
itself establish a longer coherence time.  The chiral-domain state is
experimentally reconfigurable~\cite{Dutta2026}; if the local contour pinning
assumed in the fixed-field geometric switch is achieved, erasing or resizing
the interferometer becomes a same-device control rather than a comparison
between separately fabricated samples.  Finally, heat couples directly to
neutral modes.  Localized Andreev states can contribute nonquantized subgap
heat transport and therefore cannot be dismissed a priori; the discriminant
is instead the quantized open-contact plateau together with its topology and
geometry controls.  In a fully gapped phase the non-topological bulk
quasiparticle background is exponentially suppressed for
$T\ll\Delta/\kB$.

A charged-edge converter could add the established electrical readout
of Refs.~\cite{FuKane2009,Akhmerov2009} as an independent cross-check
(Sec.~\ref{subsec:charge}), but this is intentionally separated from the core
thermal analysis because a pair of superconducting Majorana branches
is not itself a charged Dirac channel.

\paragraph{Relation to dynamical Majorana heat pumps.}
Meidan, Gur and Romito showed that a closed braiding cycle in a Majorana
Y-junction pumps a universal low-temperature heat density, and Simons, Meidan
and Romito extended that construction to the full counting statistics
~\cite{Meidan2019,Simons2020}.  The protected contribution there is the
geometric response of a genuinely multiparameter cycle in the scattering
matrix.  A vortex entering the present loop changes the discrete spin-structure
sector but is not by itself such a closed braid, and an approach--retract
one-parameter trajectory at fixed sector has no generic topologically
quantized pumping area.  We therefore keep the dynamical pumping problem
separate and use only the static elastic FCS, for which
Eq.~\eqref{eq:fcs-cgf} follows directly from the frozen transmission.

Several extensions require additional physics rather than a simple rescaling.
For $|C|>1$ the wall carries $2|C|$ co-propagating Majorana branches and the
open-contact plateau still scales with $|C|$, but generic branch mixing means
the resonant parity contrast need not equal $|C|$ times the single-channel
result.  Higher-Chern phases therefore offer a useful mode-counting staircase
in the open-contact calibration, while their interferometry should be treated with a full
multichannel scattering matrix.

Most importantly, the present two-terminal thermal loop does not measure the
$1$ versus $\psi$ fusion channel.  Recent work on an X-shaped
four-terminal Majorana interferometer shows explicitly why extra connectivity
is valuable: both vacuum and fermionic Ising fusion channels can then enter the
DC response, unlike simpler two-arm geometries~\cite{Giuliano2026}.  A
rewritable rhombohedral-graphene network could eventually supply that
connectivity, but it is a distinct extension.  Observation of the predicted signatures would instead establish a bulk-calibrated, vortex-parity-dependent
spin-structure switch of an intrinsic chiral Majorana edge, with geometry,
field and temperature controls on the same material platform.

%=======================================================================
\bibliographystyle{apsrev4-2}
\bibliography{ref_mzi_final_v2}
%=======================================================================

%=======================================================================
\appendix
\onecolumngrid
%=======================================================================
%=======================================================================

%%==========================================================================
\section{Derivation of the loop transmission}
\label{app:transmission}
%%==========================================================================

We derive Eq.~\eqref{eq:transmission}.  Let $\gamma_{\rm loop}$ denote the
chiral Majorana amplitude on the wall and $\gamma_{1,2}$ the amplitudes in the
two reservoirs.  Because the Bogoliubov--de Gennes Hamiltonian is real in the
Majorana basis, the scattering matrix obeys $S(\varepsilon)^*=S(-\varepsilon)$
and is real at $\varepsilon=0$.  Unitarity plus reality at each contact
restricts $S_i$ to $O(2)$; discarding the reflection-improper component, which
merely relabels the outgoing lead, leaves the one-parameter family of
Eq.~\eqref{eq:contact-S}.

Enumerate paths from reservoir $1$ to reservoir $2$ by the number $m$ of
complete loop traversals.  The $m=0$ path enters the loop at contact $1$
(amplitude $s_1$), propagates the arc $L_1$ (phase
$e^{i\varepsilon L_1/\hbar v}$), and exits at contact $2$ (amplitude $s_2$).
Each additional traversal remains on the loop at contact $2$ (amplitude
$c_2$), propagates $L_2$, remains on the loop at contact $1$ (amplitude
$c_1$), and propagates $L_1$; the accumulated factor per traversal is
$c_1c_2\,\sigma\,e^{i\varepsilon L/\hbar v}$, where the $Z_2$ sign appears
once per closed circuit.  Summing the geometric series, whose ratio has
modulus $c_1c_2<1$, gives Eq.~\eqref{eq:t-amplitude}, and
\begin{equation}
  T=|t_{21}|^2
   =\frac{s_1^2s_2^2}{\left|1-c_1c_2\sigma e^{i\varepsilon L/\hbar v}\right|^2}
   =\frac{(1-c_1^2)(1-c_2^2)}
         {1+c_1^2c_2^2-2c_1c_2\sigma\cos(\varepsilon L/\hbar v)} ,
\end{equation}
which is Eq.~\eqref{eq:transmission}.  The arc phase $e^{i\varepsilon
L_1/\hbar v}$ cancels from the modulus, so only the perimeter enters.
Unitarity is verified by $T+R=1$ with
$R=|r_{11}|^2$ computed from the same series.

At a resonance, $\cos(\varepsilon L/\hbar v)=\sigma$, the denominator reduces
to $(1-c_1c_2)^2$ and $T=(1-c_1^2)(1-c_2^2)/(1-c_1c_2)^2$, which equals unity
for $c_1=c_2$.  The full width at half maximum in energy is
\begin{equation}
  \Gamma \simeq \frac{2\hbar v}{L}\,\frac{1-c_1c_2}{\sqrt{c_1c_2}}
\end{equation}
for $c_1c_2\to1$, which is the quantity that controls how much heat the
Ramond zero mode can carry.

%%==========================================================================
\section{Static mixing of the two co-propagating Majorana branches}
\label{app:holonomy}
%%==========================================================================

For $|C|=1$ the $+1/-1$ wall carries two co-propagating Majorana branches.
Class D fixes their net chirality but does not forbid a local bilinear mixing.
A minimal stress test is to represent propagation around one circuit by an
$SO(2)$ holonomy $R(\phi)$ in branch space, in addition to the dynamical phase
and the vortex-parity sign.  If the two contacts are branch diagonal and have
the same scalar parameters used in the main text, diagonalizing $R(\phi)$
gives the two eigenchannel transmissions
\begin{equation}
 T_{\pm}(\varepsilon;\phi)=
 \frac{(1-c_1^2)(1-c_2^2)}
 {1+c_1^2c_2^2-2c_1c_2\sigma
 \cos\!\left(\varepsilon L/\hbar v\pm\phi\right)}.
\label{eq:holonomy-transmission}
\end{equation}
The total heat conductance is the sum of the two Landauer integrals.  The
open-contact limit is independent of $\phi$ and still gives one Dirac thermal
quantum for the $|C|=1$ wall.  At finite reflection the static holonomy shifts
the resonance comb and can reduce the vortex-parity contrast.  In particular,
for this symmetric two-branch model $\phi=\pi/2$ maps the Ramond and
Neveu--Schwarz sets of shifted resonances onto one another, so the integrated
contrast vanishes accidentally.  This is not a loss of topology: varying
$L/T$, the contact settings, or the microscopic wall configuration moves away
from the node.  The corresponding numerical stress test gives the same accidental node.  More
general energy-dependent contact and branch mixing can be treated with a
matrix-valued round-trip operator; Eq.~\eqref{eq:holonomy-transmission} is the
minimal extension needed to show which part of the scalar result is universal
and which part is device dependent.

%%==========================================================================
\section{Fusion blindness of the quadratic scattering problem}
\label{app:nogo}

A discretized Majorana representation makes the statement of
Sec.~\ref{subsec:nogo} explicit.

Discretize the closed wall into $N$ sites carrying Majorana operators
$\eta_j$, and append $n_{\rm v}$ vortex-core modes $\gamma_a$.  In the Majorana
basis the single-particle problem is the real antisymmetric matrix $A$ of
Eq.~\eqref{eq:bdg-majorana}, with entries
\begin{equation}
  A_{j,j+1}=-A_{j+1,j}=\frac{\hbar v}{2a_0},\qquad
  A_{j_a,N+a}=-A_{N+a,j_a}=\lambda_a ,
\label{eq:A-entries}
\end{equation}
$a_0$ the lattice spacing, $j_a$ the contact site of vortex $a$, and
$\lambda_a$ its hybridization amplitude, related to the width of
Eq.~\eqref{eq:hyb-width} by $\Gamma_a=\lambda_a^2/(\hbar v/2a_0)$.  For well-separated cores the direct overlap entry
$A_{N+1,N+2}$ is zero up to $\mathcal{O}(e^{-d_{12}/\xi})$.  More importantly,
the fusion occupation $\mathcal{P}=i\gamma_1\gamma_2$ labels a many-body state
of a fixed quadratic Hamiltonian; it is not an independent parameter of $A$.

Every quantity in Sec.~\ref{sec:scattering} is a functional of $A$: the
scattering matrix is $S(\varepsilon)=1-2\pi iW^\dagger(\varepsilon-
H_{\rm eff})^{-1}W$ with $H_{\rm eff}$ built from $A$ and the lead couplings
$W$, and the conductance is the thermal average of $|S_{21}|^2$.  Because changing the occupation label does not change $A$, it does not change
$S$ within this Gaussian description.  A numerical matrix check with $N=400$
and two hybridized cores verifies $A=-A^{T}$, particle-hole symmetry of the
spectrum of $iA$, and the vanishing separated-core overlap entry.

Two remarks delimit the statement.  It is a statement about Gaussian
observables, not about the physics: the fusion channel is perfectly well
defined, it is simply not visible to a probe whose observable is a functional
of $A$.  Additional contacts, arms, or hybridized cores can modify the quadratic
scattering matrix but do not by themselves make it conditional on the prepared
fusion occupation.  Representative routes to fusion-sensitive transport add
non-Gaussian $\sigma$ tunneling or a charging constraint on an island, as in
Refs.~\cite{Klocke2022,Wei2023,Giuliano2026,MicrosoftMajorana2025}.

\section{Damped-interference transmission}
\label{app:dephasing}

Expanding the resonator denominator in path pairs,
\begin{equation}
  \frac{1}{|1-ue^{i\theta}|^{2}}
  =\sum_{m,n\geq0}u^{m+n}e^{i(m-n)\theta},
  \qquad u=c_1c_2\sigma ,
\end{equation}
separates diagonal terms $m=n$, which are classical, from interference terms
$m\neq n$.  To model partial coherence, we damp off-diagonal path pairs by
$\lambda^{|m-n|}$, with $\lambda=e^{-L/\ell_\phi}$, in the spirit of a
B\"uttiker dephasing probe~\cite{Buttiker1986}.  Summing over $n$ at fixed $k=m-n$ gives
$u^{|k|}/(1-u^2)$, and the remaining geometric series in $k$ yields
\begin{equation}
  \sum_{m,n}u^{m+n}\lambda^{|m-n|}e^{i(m-n)\theta}
  =\frac{1-\lambda^2u^2}
        {(1-u^2)\left(1-2\lambda u\cos\theta+\lambda^2u^2\right)} ,
\end{equation}
which is Eq.~\eqref{eq:dephased-T} after multiplication by $T_1T_2$.  The
closed form was verified against direct summation of the double series to
nine significant figures.  Three limits check it: $\lambda=1$ returns
Eq.~\eqref{eq:transmission}; $\lambda=0$ gives $T_1T_2/(1-u^2)$, independent
of $\theta$ and of $\sigma$, so a fully dephased loop reports no parity; and
$c_i=0$ gives unity for every $\lambda$, so this dephasing model does not
spoil the open-contact plateau.

\section{Normalization of the thermal conductance}
\label{app:normalization}
%%==========================================================================

Using $-\partial f/\partial\varepsilon
=(4\kB T)^{-1}\mathrm{sech}^2(\varepsilon/2\kB T)$ and $x=\varepsilon/\kB T$,
\begin{equation}
  \int_{-\infty}^{\infty}\!\dd\varepsilon\;\varepsilon^2
  \left(-\frac{\partial f}{\partial\varepsilon}\right)
  = (\kB T)^2\int_{-\infty}^{\infty}\!\dd x\;\frac{x^2}{4}
    \mathrm{sech}^2\!\left(\frac{x}{2}\right)
  = \frac{\pi^2}{3}(\kB T)^2,
\end{equation}
using $\int_0^\infty u^2\,\mathrm{sech}^2u\,\dd u=\pi^2/12$.  Substituting in
Eq.~\eqref{eq:K-landauer} with $T(\varepsilon)\equiv1$ and $h=2\pi\hbar$ gives
$K/T=\pi^2\kB^2/6h$ per Majorana channel, which fixes the normalization used
in Eq.~\eqref{eq:K-dimensionless} and reproduces the Majorana quantum
$4.732\times10^{-13}$~W/K$^2$.  For comparison, integrating the same Fermi weight only up to a finite
energy cutoff defines the bulk kernel $\Gker$ of Ref.~\cite{Ghosh2026THT},
Eq.~\eqref{eq:Gkernel}, whose closed form
\begin{equation}
  \Gker(x)=1-\frac{12}{\pi^2}\Bigl[x^2\bigl(1-\tanh x\bigr)
  +2x\ln\!\left(1+e^{-2x}\right)-\mathrm{Li}_2\!\left(-e^{-2x}\right)\Bigr]
\end{equation}
has been verified numerically against direct quadrature.  This identity is
a normalization cross-check only; it does not imply that the transparent
resonator conductance equals $\Gker$.

%%==========================================================================
\section{Numerical evaluation and consistency checks}
\label{app:numerics}

The thermal integrals in Eq.~\eqref{eq:K-dimensionless} are evaluated on a
fixed Simpson grid over $|x|\leq60$.  Because the integrand oscillates with
period $2\pi\alpha$, the grid is chosen adaptively in density, with at least
$40$ points per oscillation and a minimum of $4001$ points.  This procedure is
stable down to $\alpha=0.02$ and agrees with adaptive quadrature to six digits
where the latter converges.  The same resolved-grid integration is used for
the contrast, hybridization and dephasing scans.

Internal checks include: transparent contacts returning one Majorana quantum
per branch to $10^{-8}$; the closed form of $\Gker$ agreeing with direct
quadrature to $10^{-12}$; sector degeneracy in the thermally smeared
$\alpha\ll1$ limit; the resonant side-coupled-Majorana phase approaching
$\pi$ at zero energy and zero far from resonance; antisymmetry and
particle-hole symmetry of the discretized Majorana BdG matrix; and the
dephasing expression reducing to the coherent result at $\lambda=1$ and to a
parity-blind classical ladder at $\lambda=0$.  Direct summation of the
damped multiple-traversal series agrees with the closed form to nine
significant figures at representative parameters.  The full-counting-statistics
calculation independently verifies the transparent-channel limit, the
linear-response reduction of Eq.~\eqref{eq:fcs-mean} to
Eq.~\eqref{eq:K-dimensionless}, the equilibrium identity
Eq.~\eqref{eq:fcs-fdt}, and the Gallavotti--Cohen symmetry of
Eq.~\eqref{eq:fcs-cgf}.

The field trace in Fig.~\ref{fig:squarewave} is intentionally a reference
waveform: it uses one-by-one vortex counting
$n_v=\lfloor B/\Delta B+1/2\rfloor$ with no model of pinning, entry barriers or
hysteresis.  Consequently, the equal field spacing in that figure is not used
as an experimental prediction; the observable is the conductance switch
correlated with a controlled change of enclosed vortex parity.

%%==========================================================================
\section{Parameter dependence of the edge velocity}
\label{app:velocity}
%%==========================================================================

The principal material scale entering the finite-size response is
$v=\Delta_0/\hbar k_F$.  Table~\ref{tab:velocity} tabulates it, together with
the derived scales, over the illustrative gap ratios used in
Ref.~\cite{Ghosh2026THT}.  The value $k_F=0.29$~nm$^{-1}$ is quoted directly from the degeneracy-two
R8G Landau fan of Ref.~\cite{Okounkova2026} and is used only as a representative,
conservative momentum scale; the R4G values discussed in
Ref.~\cite{Ghosh2026THT} are slightly smaller and would increase $v$ at fixed $\Delta_0$.  The reduced finite-size scale $\hbar v/L$ scales linearly with $\Delta_0$,
so a device with a larger gap ratio permits a
larger loop at fixed $\alpha$ and thereby widens the overlap between the
high-contrast region and the field-sweep guideline of Eq.~\eqref{eq:L-min}.  Direct spectroscopic determination of $\Delta_0$ would therefore reduce the
dominant uncertainty in this scale estimate; the numerical benchmarks here do
not assume such a determination.  This caution is reinforced by direct
Meissner imaging in R3G/WSe$_2$, where $\rho_s(T)$ is inconsistent with an
isotropic BCS form and is instead well described over the measured range by
$\rho_s(T)=\rho_s^0[1-(T/T_c)^n]$ with a typical $n\simeq1.9$
~\cite{ZhangMeissner2026}.  The same study finds $\xi\lesssim80$~nm and
$\ell_{\rm MF}\gtrsim200$~nm, independently supporting a clean-limit RHG
regime.  Because that device is R3G/WSe$_2$, neither its gap shape nor its
length scales are imported into the R4G/R5G numerical benchmark; they only
reinforce treating $\Delta_0$ as an experimental input rather than a fixed
BCS ratio.

\begin{table}[htbp]
  \centering
  \caption{Edge velocity and derived scales at $k_F=0.29$~nm$^{-1}$ and
    $T_c=300$~mK.  $\lambda_T$ and the reduced finite-size scale
    $\hbar v/(L\kB)$ are quoted at $12$~mK and $L=1\,\mu$m.}
  \label{tab:velocity}
  \begin{ruledtabular}
  \begin{tabular}{ccccc}
    $2\Delta_0/\kB T_c$ & $\Delta_0$ (meV) & $v$ (m/s)
      & $\lambda_T$ ($\mu$m) & $\hbar v/L\kB$ (mK) \\
    \hline
    10 & 0.129 & \phantom{0}677 & 2.71 & \phantom{0}5.17 \\
    20 & 0.259 & 1354 & 5.42 & 10.34 \\
    30 & 0.388 & 2032 & 8.12 & 15.52 \\
  \end{tabular}
  \end{ruledtabular}
\end{table}

%%==========================================================================
\section{Domain of validity}
\label{app:assumptions}
%%==========================================================================

The calculation separates established topological inputs from device-level
assumptions and model-dependent scale estimates.  In particular, applying the
$+C/-C$ wall construction to the reconfigurable RHG domains assumes that the
opposite chiral domains also carry opposite BdG Chern numbers; present domain
imaging establishes chirality reversal but not this topological identification.

\emph{Established elsewhere and taken as given.}  Bulk-boundary correspondence
for class D~\cite{Read2000,StoneRoy2004,TeoKane2010}; the protected vortex
zero-mode parity for odd $\CBdG$~\cite{Volovik1999,Read2000,Ivanov2001}; the
Ising fusion rule and the vortex-dependent spin structure of a closed Majorana
edge~\cite{MooreRead1991,Nayak2008,Kitaev2006,DiFrancesco1997,Alicea2012}; the
occupied-vortex rule and the gappedness criterion
Eq.~\eqref{eq:gapped}~\cite{LeNir2026,Ghosh2026THT}; and the quantization of
thermal conductance per chiral channel~\cite{Kane1997,Jezouin2013,Banerjee2017}.

\emph{Device-level assumptions.}  That a sufficiently
smooth closed chiral-domain wall can be pinned and reconfigured on the
submicron-to-micron scale; that the adjacent bulk satisfies
Eq.~\eqref{eq:gapped}; that the contacts can be tuned into the weak-coupling
resonator regime; and that branch mixing is either small or can be calibrated
through its $L/T$ dependence.  The fixed-field geometric switch further requires two matched
contours that place the same pinned vortex outside or inside the loop at fixed
field.  Dutta et al. demonstrate deterministic domain reconfiguration
~\cite{Dutta2026}, but not yet this local contour control.  The charge-based cross-check has the
additional, separate requirement of a genuine charged chiral edge and is not
used in the thermal analysis.

\emph{Quantified sensitivities.}  Within the phenomenological dephasing
model, the multiple-traversal sum is exact and gives half contrast at
$\ell_\phi\simeq2.3L$ for the representative point
$\alpha=1.44$, $c_1=c_2=0.93$.  Edge-vortex hybridization provides an
intrinsic false-positive channel with independent temperature and geometry
discriminators (Sec.~\ref{subsec:artefact}).  The actual $\ell_\phi$ of an RHG
chiral Majorana wall remains an experimental input.

\emph{Not assumed.}  No claim is made about braiding or about measurement of the $1$ versus
$\psi$ fusion channel.  No quasiparticle-poisoning time is used
to predict the vortex-parity signal, because the geometric vortex count is not
the same variable as localized fermion occupation.  We also do not assume a
perfectly periodic sequence of vortex-entry fields.  Coherent propagation and
sufficiently weak inelastic equilibration around the resonator are required;
these are device-level conditions to be tested experimentally rather than
consequences of topology.

\end{document}